\documentstyle [epsf]{article}
\begin{document}
\def\reals{{\bf R}}
\def\ints{{\bf Z}}
\def\z2{{\bf Z}_2}
\def\sgn{{\rm sgn}}
\def\cc{{\rm c.c.}}
\def\eq{\begin{displaymath}}
\def\endeq{\end{displaymath}}
\def\eqno{\begin{eqnarray}}
\def\endeqno{\end{eqnarray}}
\def\eqalign{\begin{eqnarray*}}
\def\endalign{\end{eqnarray*}}
\def\k{{\bf k}}
\def\r{{\bf r}}
\def\W{{\bf W}}
\def\whex{w_{\rm hex}}
\title{\bf Simple and Superlattice Turing Patterns in Reaction-Diffusion
	Systems: Bifurcation, Bistability, and Parameter Collapse
}
\author{Stephen L. Judd and Mary Silber\\
{\it Department of Engineering Sciences and Applied Mathematics}\\
{\it Northwestern University} \\
{\it Evanston, IL 60208 USA}}
\maketitle

\begin{abstract}
We use equivariant bifurcation theory to investigate pattern selection
at the onset of a Turing instability in a general two-component
reaction-diffusion system. The analysis is restricted to patterns that
periodically tile the plane in either a square or hexagonal fashion.
Both simple periodic patterns (stripes, squares, hexagons, and rhombs)
and ``superlattice'' patterns are considered. The latter correspond to
patterns that have structure on two disparate length scales; the short
length scale is dictated by the critical wavenumber from linear
theory, while the periodicity of the pattern is on a larger scale.
Analytic expressions for the coefficients of the leading nonlinear terms
in the bifurcation equations
are computed from the general reaction-diffusion system using
perturbation theory. 
We show that, no matter how complicated the reaction kinetics might be,
the nonlinear reaction terms enter the analysis through just four
parameters.  Moreover, for hexagonal problems, all patterns
bifurcate unstably unless a particular degeneracy condition is
satisfied, and
at this degeneracy we find that the number of effective
system parameters drops to two, allowing a complete characterization
of the possible bifurcation results at this degeneracy. For example,
we find that rhombs, squares and superlattice patterns always
bifurcate unstably.  
We apply these general results to some specific model equations,
including the Lengyel-Epstein CIMA model, to investigate the
relative stability of patterns as a function of system parameters,
and to numerically test the analytical predictions.
\end{abstract}

\section{Introduction}

Regular patterns arise in a wide variety of physical, chemical and
biological systems that are driven from equilibrium. The origin of
these spatial patterns can often be traced to a symmetry-breaking
instability of a spatially-homogeneous state.  Such an instability
mechanism was proposed by Alan Turing in 1952~\cite{ref:turing} for
pattern formation in chemical systems featuring both reaction and
diffusion.  The key insight of Turing was that diffusion, which is
usually thought of as a stabilizing mechanism, can actually act as a
destabilizing mechanism of the spatially-uniform state.  This is the
case, for example, in a class of two-component activator-inhibitor
systems, provided the inhibitor diffuses more rapidly than the
activator. This competition between long-range inhibition and
short-range activation can lead to spatially inhomogeneous steady
states called Turing patterns~\cite{review}.  Experimental evidence
of Turing patterns was discovered only recently\cite{ref:castets},
and a number of experiments
on the Turing instability have now been carried out on the
chlorite-iodide-malonic acid (CIMA) reaction in a
gel~\cite{ref:castets, ref:swinney1, ref:swinney3}.  These
experiments have demonstrated the existence and stability of a
variety of regular patterns, including stripes, hexagons, and rhombs.
At the same time, a two variable model of the CIMA reaction has been
proposed by Lengyel and Epstein \cite{ref:lengep1}.

Equivariant bifurcation theory~\cite{GSS} is a powerful tool for
analyzing the nonlinear evolution of symmetry-breaking instabilities
in pattern-forming systems.  This approach determines the generic
behaviors associated with bifurcation problems that are equivariant
with respect to a given group of linear transformations.  In
particular, the form of the bifurcation problem is restricted by
symmetry considerations, limiting the behavior to some finite set of
possibilities.  The problem of determining which type of behavior
occurs in a given physical system then reduces to one of
computing the coefficients of specific nonlinear terms in the bifurcation
problem. In this paper we perform this computation for a general
two-component reaction-diffusion system in a neighborhood of a Turing
bifurcation. We consider two different symmetry groups for the
problem, one associated with Turing patterns that tile the plane in a
square lattice, and the other associated with patterns that tile the
plane in a hexagonal lattice. The symmetry groups are ${\rm D}_4\dot +
{\rm T}^2$ and ${\rm D}_6 \dot + {\rm T}^2$, respectively, where ${\rm
D}_n$ ($n=4,6$) is the dihedral group of symmetries of the fundamental
tile, and ${\rm T}^2$ is the two-torus of translation symmetries
associated with the doubly-periodic solutions.  The fundamental
irreducible representations of these groups, which are four- and
six-dimensional, lead respectively to bifurcation problems that
address competition between stripes and simple squares, and between
stripes and simple hexagons. The higher dimensional irreducible
representations, which are eight- and twelve-dimensional,
lead to bifurcation problems that
address competition between these simple Turing patterns and a class
of patterns that are spatially-periodic on a larger scale than the
simple patterns, but still arise in a primary bifurcation from the
homogeneous state at the Turing bifurcation point~\cite{ref:dionne1,
ref:mary, ref:annemary, ref:sp}. An example of such a ``superlattice''
pattern was recently observed in an experiment on parametrically
excited surface waves~\cite{ref:gollub}.  Finally, the
higher-dimensional group representations also allow us to investigate
the relative stability properties of rhombs and squares or 
rhombs and hexagons.

A number of earlier studies have used bifurcation theory to
investigate Turing pattern formation for reaction-diffusion systems.
For instance, Othmer~\cite{ref:othmer} described a group theoretic
approach for analyzing the Turing bifurcation nearly twenty years
ago, in 1979. Recently
Callahan and Knobloch~\cite{ref:calknob1} derived and analyzed the
general equivariant bifurcation problems for three-dimensional Turing
patterns having the spatial periodicity of the simple cubic,
face-centered cubic and body-centered cubic lattices. They applied
their general results to the Brusselator and the Lengyel-Epstein CIMA
reaction-diffusion models \cite{ref:calknob2}. Gunaratne, {\it et
al.}~\cite{ref:gemunu} developed an approach to pattern formation
problems based on a model equation akin to Newell-Whitehead-Segel, but
with full rotational symmetry.  They used this model to investigate
bistability of hexagons and rhombs, and tested their predictions 
against laboratory experiments.

In addition to these general bifurcation studies of Turing pattern
formation, a number of specific reaction-diffusion models have been
investigated analytically and numerically. Dufiet and
Boissonade~\cite{ref:dufiet} investigated numerically the formation of
two-dimensional Turing patterns for the Schnackenberg
reaction-diffusion system.  The formation of three-dimensional Turing
structures has been investigated for the Brusselator by DeWit, {\it et
al.}~\cite{ref:dewit}. Turing patterns for the two-variable
reaction-kinetics based model of the CIMA reaction, proposed by
Lengyel and Epstein~\cite{ref:lengep1}, have been investigated using
bifurcation theory~\cite{ref:rovinsky}, as well as
numerically~\cite{ref:jensen1}.

This paper extends existing bifurcation studies of Turing patterns in
reaction-diffusion systems in two primary ways. The first, as
discussed above, is that we consider the higher dimensional (8 and 12,
respectively) irreducible group representations associated with the
square and hexagonal lattice tilings. This corresponds to choosing a
fundamental tile for the periodic patterns large enough to admit many
copies of the simple Turing patterns -- stripes, hexagons, squares,
and rhombs -- as well as new solutions in the form of ``superlattice''
patterns. In this way we extend the standard stripes {\it vs.}
hexagons stability analysis, while also determining conditions under
which other more exotic Turing patterns might arise.

The second way in which our analysis differs from earlier studies is
that it does not restrict attention to a specific model ({\it e.g.}
Schnackenberg, Brusselator, Lengyel-Epstein CIMA model, {\it
etc.}). The coefficients in the bifurcation equations are derived for
a general two-component reaction diffusion system. This allows us to
make a number of general statements about Turing patterns in these
systems.  For
example, we show that the bifurcation results depend on four effective
parameters that may be simply computed from the nonlinear terms in the
reaction-kinetics.  Moreover, in the case of the standard degenerate
hexagonal bifurcation problem, we show that the details of the
reaction-kinetics enter the analysis through just {\it two}
parameters. Even more surprising is the result that at this degenerate
point all angle-dependence, as associated with rhombic patterns, drops
out of the corresponding bifurcation problems; in this case we show
that all rhombic patterns are unstable to rolls, even when hexagons are
the preferred planform. This is in contrast to the general
stability analysis of Gunaratne {\it et
al.}~\cite{ref:gemunu}, which suggest that rhombs with angle close to
$\pi/3$ should be stable when hexagons are stable. Thus our analysis
indicates that two-component reaction-diffusion models may have a more
rigid mathematical structure than anticipated.

Our paper is organized as follows: Section~\ref{sec:formulation}
reviews the necessary conditions for a Turing instability in a
two-component reaction-diffusion system. It also provides the
necessary background on the mathematical framework for the bifurcation
analysis. Section~\ref{sec:reduc} calculates the bifurcation equation
coefficients using perturbation methods, with the final results
summarized in an Appendix. Section~\ref{sec:collapse} describes the
concept of parameter collapse, where multiple system parameters
collapse into a small number of effective parameters.
Special attention is given to the degenerate
bifurcation problem on the hexagonal lattice, where we show that the details
of the reaction-kinetics enter the analysis in this case through only
two effective parameters.  The resulting predictions are then
then tested numerically;  specifically, we present
an example of tri-stability between stripes, simple hexagons, and
super hexagons in a neighborhood of the degenerate bifurcation
problem.  Section~\ref{sec:applications} applies the general
results to a model of the CIMA reaction proposed by Lengyel and
Epstein~\cite{ref:lengep1} and a model demonstrating super squares.  
Finally, we suggest some directions for
future work and summarize the key points of the paper in the conclusions
section.

\bigskip

\section{Formulation of Bifurcation Problem} \label{sec:formulation}

\subsection{Linearized Problem}

We begin by summarizing the conditions for a Turing instability to occur
as some parameter is varied in the reaction-diffusion system
\begin{eqnarray}
\label{eq:full}
        u_t &= &\nabla^2 u + f(u,v)\nonumber\\
        v_t &= &K\nabla^2 v + g(u,v),
        \quad \nabla^2 = \partial_{xx} + \partial_{yy}\ .
\end{eqnarray}
We assume that the species diffuse at different rates and that $v$
measures the concentration of the more rapidly diffusing species. Thus
the ratio of diffusion coefficients is $K>1$. The Turing instability
first occurs as a symmetry-breaking steady state bifurcation of a
spatially-uniform steady state of (\ref{eq:full}); associated with the
instability is a critical wave number $q_c\ne 0$, which is determined
below.

Without loss of generality, we take the spatially-uniform state
to be $u=v=0$.
We expand $f(u,v)$ and $g(u,v)$ about this state as
\begin{eqnarray}
f(u,v)& =& au + bv + \ldots\nonumber\\
g(u,v)& = & cu + dv + \ldots\nonumber
\end{eqnarray}
yielding the linearized reaction-diffusion system 
\eqno
\label{eq:linear}
	\left(\matrix{u_t\cr v_t\cr}\right) = 
		\left(\matrix{ a + \nabla^2& b\cr c &d+K\nabla^2\cr}\right)
		\left(\matrix{u \cr v\cr}\right).
\endeqno
The stability of the spatially uniform solution is determined
by substituting
$$\left( \matrix{u\cr v}\right) =
	{\bf \xi} e^{iqx}e^{\sigma t}$$
into~(\ref{eq:linear}) to obtain the eigenvalue problem
\eqno
\label{eq:lq}
\sigma(q) {\bf \xi} = L(q){\bf \xi};
\quad
	L(q)\equiv	\left(\matrix{ a - q^2& b\cr c
&d-Kq^2}\right)\ .
\endeqno
The eigenvalues $\sigma_1(q),\ \sigma_2(q)$ depend on the parameters
$a,b,c,d,K$, as well as the wavenumber of the perturbation $q$. A
Turing bifurcation occurs for parameter values where there is a zero
eigenvalue for some $q=q_c\ne 0$ with all other modes being
damped, {\it i.e.},
$$\sigma_1(q_c) < \sigma_2(q_c) = 0,$$
$${\rm Re}\big(\sigma_j(q)\big) < 0, \quad j=1,2,\quad {\rm for\ all}\
q\ne q_c.$$
These inequalities imply that
\eqno
	{\rm Tr}\left(L(q)\right) &<& 0,\quad \forall q,\label{eq:lin1}\\
	{\rm Det}(L(q)) &>& 0,\quad q\ne q_c,
\nonumber
\endeqno
at the bifurcation point.
The critical wave number $q_c$ is given by
\eqno
	&{\rm Det}(L(q_c)) =0.& \label{eq:lin3}
\endeqno
Since ${\rm Det}(L(q))$ is quadratic in $q^2$ and concave up,
the second of conditions~(\ref{eq:lin1}) is assured if $q_c$ is
a minimum of ${\rm Det}(L(q))$:
\eqno
 \left. {\partial {\rm Det}(L(q))\over \partial q} \right|_{q=q_c} = 0.
 \label{eq:lin4}
\endeqno
Thus the parameters at the Turing bifurcation 
($a=a_c,\ b=b_c,\ c=c_c,\ d=d_c,\ K=K_c,\ q=q_c$) satisfy 
\eqno
q_c^2 = {K_ca_c + d_c\over 2K_c}>0,\label{eq:qc}
\endeqno
\eqno
(K_ca_c-d_c)^2 + 4K_cb_cc_c = 0, \label{eq:critparms}
\endeqno
and must also satisfy the inequalities
\eqno
  K_ca_c+d_c&>&0,\nonumber\\
  b_cc_c&< &0, \label{eq:negcond}\\
  a_c+d_c &< &0,\nonumber\\
  a_cd_c-b_cc_c &> &0,\nonumber
\endeqno
to ensure that $q_c^2>0$ and that
perturbations at $q=0$ decay.
From the initial assumption that
$K>1$, it follows that
$a_c>0$ and $d_c<0$,
which for $b_c<0,\ c_c>0$ is consistent with the interpretation of $u$ as
the concentration of the activator and $v$ as the concentration of the
inhibitor in~(\ref{eq:full}).

\subsection{Spatially-Periodic Solutions of the Nonlinear Problem}
\label{sec:lattice}

For the two-dimensional spatially-unbounded problem there are
infinitely many neutrally stable Fourier modes at the bifurcation
point: all Fourier modes associated with wave vectors on the critical
circle $|{\bf q}|=q_c$ are neutrally stable.  In the remainder of the
paper we focus on spatially doubly-periodic solutions
of~(\ref{eq:full}): $$u({\bf r},t)=u({\bf r}+{\bf d}_1,t)=u({\bf
r}+{\bf d}_2,t),\quad v({\bf r},t)=v({\bf r}+{\bf d}_1,t)=v({\bf
r}+{\bf d}_2,t), $$ where ${\bf r}\equiv(x,y)$ and ${\bf d}_1,\ {\bf
d}_2$ are linearly independent vectors in ${\bf R}^2$.  We further
restrict to the cases where $|{\bf d}_1|=|{\bf d}_2|$. We then write
each field $u,v$ in terms of its Fourier series, {\it e.g.}
\begin{eqnarray}
\label{eq:fourier}
  u({\bf r},t) &= &\sum_{j,k\in {\bf Z}}^{} \hat u_{j,k}(t) 
	e^{i q (j\widehat {\bf k}_1 
	+ k \widehat {\bf k}_2) \cdot {\bf r}}\ ,\quad
  	\hat{u}_{-j,-k}=\hat u^*_{j,k},
\end{eqnarray}
where $\widehat {\bf k}_1$ and $\widehat {\bf k}_2$ are unit vectors that
satisfy 
$$q\widehat{\bf k}_i\cdot {\bf d}_j=2\pi \delta_{ij}$$ 
for an appropriate choice of $q$. Since the Fourier transform of
spatially-periodic solutions is discrete, only a finite number of
Fourier modes can have wave vectors ${\bf q}$
that lie on the critical circle $|{\bf q}|=q_c$ at the bifurcation point;
all other modes will be linearly damped.
Thus, in this setting, finite-dimensional bifurcation theory may be
applied in a neighborhood of the Turing bifurcation point.

\begin{figure}
  \epsfbox{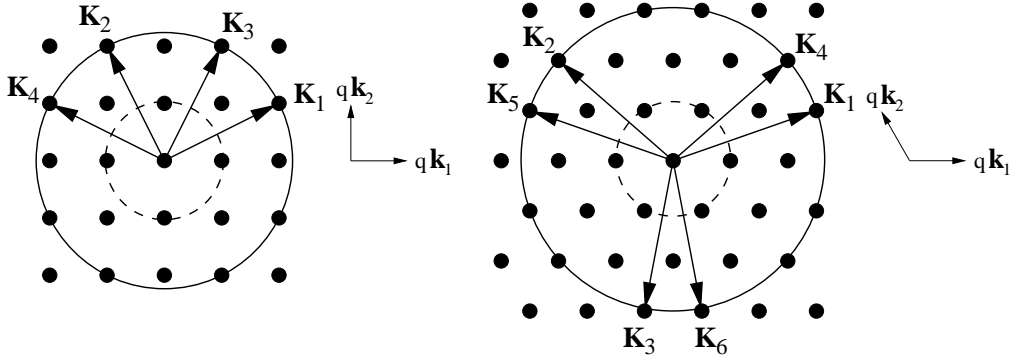}
  \caption{Square and hexagonal lattices 
	in Fourier-space plotted together with two examples of
	critical circles of radius $q_c$.
	The dashed circles corresponds to $q_c=q$ in
(\protect\ref{eq:qsmn}) and (\protect\ref{eq:qhmn}),
{\it i.e.} $(m,n)=(1,0)$; the solid
	circle corresponds to $(m,n)=(2,1)$ on the square lattice
	and $(m,n)=(3,1)$ on the hexagonal lattice.
	Note that the solid
	circle has twice as many critical modes as the dashed circle.
  }
  \label{fig:lattice}
\end{figure}

For Turing patterns that tile the plane in a square lattice the unit
vectors $\widehat{\bf k}_1$ and $\widehat{\bf k}_2$ are separated by an
angle of $\pi/2$, whereas in the hexagonal case they are separated by
$2\pi/3$; all angles that are not a multiple of $\pi/3$ or
$\pi/2$ correspond to rhombic lattices. The parameter $q$
in~(\ref{eq:fourier}), which is determined by the imposed period, also
dictates the the spacing of the lattice points in the Fourier
transform, and may in turn be used to determine which lattice points
(if any) lie on the critical circle.
Of particular interest here is the ratio of this scale to the natural
scale of the problem $q_c$. For instance, if $q/q_c=1$ then on the
square lattice there are four lattice points which lie on the critical
circle: $(j,k)=(\pm 1,0)$, $(0,\pm 1)$ in (\ref{eq:fourier}),
corresponding to Fourier modes $$e^{iq_c x}, \quad e^{-iq_c x},
\quad e^{i q_c y}, \quad e^{-iq_c y}.$$ If, however, a tighter spacing
with $q/q_c=1/\sqrt{5}$ is used then there are eight square-lattice
modes which lie on the circle: $(j,k)=(2,\pm 1)$, $(-2,\pm 1)$,
$(1,\pm 2)$ and $(-1,\pm 2)$. Examples for these cases, as well as
analogous ones for the hexagonal lattice, are depicted in
Figure~\ref{fig:lattice}.

In this paper we consider a discrete set of values of $q$,
parameterized by an integer pair $(m,n)$, which admit wave
vectors in~(\ref{eq:fourier}) with length $q_c$, {\it i.e.}, for which
$|q(m\widehat{\bf k}_1+n\widehat{\bf k}_2)|=q_c$. 
Specifically, on the square lattice the values of $q$ are related to
$(m,n)$ by 
\eqno
\label{eq:qsmn}
  q^2 = q_c^2/(m^2 + n^2),
\endeqno
and on the hexagonal lattice by 
\eqno
\label{eq:qhmn}
  q^2 = q_c^2/(m^2 + n^2 - mn)\ .
\endeqno

In a neighborhood of the bifurcation point,
$(a, b, c, d, K)=(a_c, b_c, c_c, d_c, K_c)$, we apply our analysis
separately to the cases where there are eight critical Fourier modes
on the square lattice and twelve on the hexagonal lattice (see
Figure~\ref{fig:lattice}).
Specifically, on the square lattice, we investigate primary steady solution
branches of the form 
\eqno
\label{eq:sqcase}
	z_1 e^{iq {\bf K}_1\cdot {\bf r}} + z_2 e^{iq {\bf K}_2\cdot {\bf r}} 
	+ z_3 e^{iq{\bf K}_3\cdot {\bf r}} + z_4 e^{iq{\bf K}_4\cdot {\bf r}}
	+ \cc + \cdots
\endeqno 
where $z_1, \ldots, z_4$ are the 
complex-valued amplitudes of the critical modes,
$\cdots$ represents the linearly-damped Fourier modes
that are slaved to the critical modes,
 and
${\bf K}_1=(m,n)$, ${\bf K}_2=(-n,m)$, 
${\bf K}_3= (n,m)$, and ${\bf K}_4=(-m,n)$.
Note that, while
${\bf K}_1\cdot {\bf K}_2={\bf K}_3\cdot{\bf K}_4=0$, the angle
between ${\bf K}_1$ and ${\bf K}_3$ depends on the integer pair
$(m,n)$. Similarly, on the hexagonal lattice,
\eqno
\label{eq:hexcase}
\sum_{n=1}^{6} z_n e^{iq{\bf K}_n\cdot {\bf r}} + \cc + \cdots
\endeqno
where ${\bf K}_1=m(1,0) + n(-1/2, \sqrt{3}/2)$ and 
${\bf K}_4=m(1,0) + (m-n)(-1/2, \sqrt{3}/2)$.
${\bf K}_2$ and
${\bf K}_3$ are obtained by rotating ${\bf K}_1$ by $\pm 2\pi/3$, and
${\bf K}_5$ and ${\bf K}_6$ are obtained by rotating ${\bf K}_4$ by
$\pm 2\pi/3$ (see Figure~\ref{fig:lattice}). 
Note that the angle between ${\bf K}_1$ and ${\bf K}_4$ again
depends on $(m,n)$.

Dionne and Golubitsky \cite{ref:dionne1} used the equivariant
branching lemma \cite{GSS} to prove that for the square lattice
cases~(\ref{eq:sqcase}), the following types of steady solutions arise
in pitchfork bifurcations from the spatially-uniform state:
\begin{itemize}
  \item Stripes: ${\bf z}\equiv (z_1,z_2,z_3,z_4) = (x,0,0,0)$
  \item Simple squares: ${\bf z}=(x,x,0,0)$
  \item Rhombs: ${\bf z}=(x,0,x,0)$ and $(x,0,0,x)$
  \item Super squares: ${\bf z}=(x,x,x,x)$
  \item Anti-squares: ${\bf z}=(x,x,-x,-x)$, 
\end{itemize}
where in each case $x$ is a real amplitude.  (Examples of super and 
anti-squares are given in Section~\ref{sec:supsq}, 
Figure~\ref{fig:sasexample}).  While
the striped and simple square patterns are essentially the same for each
$m$ and $n$, the solutions in the form of rhombs,
super squares and anti-squares differ for each co-prime $m$ and
$n$ since the angle between ${\bf K}_1$ and ${\bf K}_3$ (and
${\bf K}_1$ and ${\bf K}_4$) depends
on $m$ and $n$.  The general form of the
amplitude equations describing the branching and relative stability of
the square lattice patterns is derived in~\cite{ref:mary}. To cubic
order the bifurcation equations take the simple form
\eqno
\dot z_1 = \mu z_1 + z_1(a_1 |z_1|^2 + a_2|z_2|^2 + a_3 |z_3|^2
	+ a_4 |z_4|^2) + \cdots \label{eq:sqeq}
\endeqno
with similar expressions for $\dot z_2$, $\dot z_3$, $\dot z_4$ 
and their conjugates.  Here
$\mu$ is the bifurcation parameter and $a_1, \ldots, a_4$ are real.
Linearizing~(\ref{eq:sqeq}) about a particular solution generates a
set of eigenvalues which determines the stability of that solution
with respect to perturbations on the lattice.  The signs
of these eigenvalues
are given in Table~\ref{tab:evsq} in terms of $a_1, \ldots, a_4$; 
a positive eigenvalue indicates
instability. Note that higher order terms are
necessary to determine the relative stability of the super square and
anti-square patterns~\cite{ref:mary}. In the next section we compute
the coefficients $a_1,\ldots,a_4$ needed to evaluate the signs of the
eigenvalues
in terms of the angle between ${\bf K}_1$ and ${\bf K}_3$
and in terms of parameters in the general reaction diffusion
system~(\ref{eq:full}).

\begin{table}
\centering
\caption{Signs of eigenvalues for primary bifurcation branches on the
square lattice; $a_1,\ldots,a_4$ are coefficients in the
bifurcation equation~(\protect\ref{eq:sqeq}).  
See~\protect\cite{ref:mary} for more details.}
\label{tab:evsq}
\begin{tabular}{|c|c|}
\hline
Planform & Signs of non-zero eigenvalues \\
\hline 
 & \\
Stripes  &  $\sgn(a_1), \quad \sgn(a_2 - a_1), 

        \quad \sgn(a_3 - a_1), \quad \sgn(a_4 - a_1)$     \\ 
${\bf z}=(x,0,0,0)$  & \\
& \\
\hline
 & \\
Simple Squares
       &  $\sgn(a_1 + a_2), \quad  \sgn(a_1 - a_2), \quad
          \sgn(a_3 + a_4 - a_1 - a_2)$ \\
${\bf z}=(x,x,0,0)$ & \\
&\\
\hline
 & \\
Rhombs
       &  $\sgn(a_1 + a_3), \quad \sgn(a_1 - a_3), \quad
         \sgn(a_2 + a_4 - a_1 - a_3)$  \\
${\bf z}=(x,0,x,0)$& \\
&\\
\hline
& \\
Rhombs
       &  $\sgn(a_1 + a_4), \quad \sgn(a_1 - a_4), \quad     
         \sgn(a_2 + a_3 - a_1 - a_4)$  \\ 
${\bf z}=(x,0,0,x)$& \\
&\\
\hline
 & \\ 
Super Squares 
       &  $ \sgn(a_1 + a_2 + a_3 + a_4), \quad
         \sgn(a_1 + a_2 - a_3 - a_4)$   \\          
${\bf z}=(x,x,x,x)$&  $\sgn(a_1 - a_2 + a_3 - a_4), \quad
         \sgn(a_1 - a_2 - a_3 + a_4)$   \\
 & $\sgn(\zeta_0)$, where 
	$\zeta_0={\rm O}\bigl(x^{2(m+n-1)}\bigr)$\\
&\\
\hline
 & \\ 
Anti--Squares
       & same as super squares, except $\zeta_0\to -\zeta_0$\\ 
${\bf z}=(x,x,-x,-x)$
       & \\
 & \\
\hline
\end{tabular}
\end{table}


Dionne and Golubitsky \cite{ref:dionne1} also considered the hexagonal
lattice problem and used group theoretic methods to determine that
there are primary branches of the form
\begin{itemize}
  \item Stripes: ${\bf z}\equiv(z_1,z_2,z_3,z_4,z_5,z_6) = (x,0,0,0,0,0)$
  \item Simple hexagons: ${\bf z}=(x,x,x,0,0,0)$
  \item Rhombs: ${\bf z}=(x,0,0,x,0,0)$, $(x,0,0,0,x,0)$, and $(x,0,0,0,0,x)$
  \item Super hexagons: ${\bf z}=(x,x,x,x,x,x)\ .$
\end{itemize}
Recently, Silber and Proctor \cite{ref:sp} showed that there is an
additional primary solution branch that has triangular symmetry:
\begin{itemize}
\item Super triangles: ${\bf z}=(z,z,z,z,z,z)\ \quad z\in {\bf C},$
\end{itemize}
where, near the bifurcation point, the argument of the complex
amplitude $z$ is $\sim {\pi\over 3},{2\pi\over 3}$.  The stripes and simple
hexagons are the same for each integer pair $(m,n)$, while
the other solutions change with the values of $m$ and
$n$. The general form of the bifurcation equations describing the
evolution of the critical modes $z_1,\ldots,z_6$ is derived 
in~\cite{ref:mary}. The cubic truncation is
\eqno
\dot z_1 & = &\mu z_1 + \gamma z_2^* z_3^* + z_1(
	b_1 |z_1|^2 + b_2(|z_2|^2 + |z_3|^2) \cr
	&&\qquad\qquad\qquad\qquad + b_4 |z_4|^2 + b_5 |z_5|^2 + b_6 |z_6|^2)
	+ \cdots \label{eq:hexeq}
\endeqno
with similar expressions for $\dot z_2,\ldots,\dot z_6$. While the
stripes and rhombs solutions arise from the uniform state in a
pitchfork bifurcation, all other solutions
bifurcate transcritically.  The relative stability of the primary
patterns are summarized from
\cite{ref:mary} in Table~\ref{tab:evhex}.
 Note that unless the 
quadratic coefficient $\gamma$ is zero, all branches bifurcate
unstably. Thus we will consider a degenerate bifurcation problem for
which $\gamma \approx 0$. Also note that, as in the case of super squares 
and anti-squares, the relative stability of the super hexagons
and super triangles is determined by terms higher than cubic in
(\ref{eq:hexeq}) \cite{ref:sp}.

\begin{table}
\centering
\caption{Branching equations and signs of eigenvalues for 
primary bifurcation branches on the hexagonal lattice; $\gamma,
b_1,\dots,b_6$ are coefficients in the bifurcation equation
(\protect\ref{eq:hexeq}).  See \protect\cite{ref:mary, ref:annemary} 
for more details.}
\label{tab:evhex}
\begin{tabular}{|c|c|}
\hline
Planform and branching equation	     &  Signs of non-zero eigenvalues \\ 

\hline
& \\ 
Stripes
       & $\sgn(b_1)$,
\\ 
${\bf z}=(x,0,0,0,0,0)$ &
 	$\sgn(\gamma x+(b_2-b_1)x^2), \quad \sgn(-\gamma x+(b_2-b_1)x^2)$,\\

$0=\mu x +b_1 x^3+{\rm O}(x^5)$ &            
$\sgn(b_4 - b_1), \quad \sgn(b_5 - b_1), \quad \sgn(b_6 - b_1)$\\
& \\

\hline
& \\ 
Simple Hexagons
       & $ \sgn(\gamma x+2(b_1+2b_2)x^2), \quad 
\sgn(-\gamma x+(b_1-b_2)x^2$)               \\ 
${\bf z}=(x,x,x,0,0,0)$ &
 $ \sgn(-\gamma x+(b_4+b_5+b_6-b_1-2b_2)x^2)$  \\ 
$0=\mu x +\gamma x^2 + (b_1 + 2b_2) x^3 + {\rm O}(x^4)$
       & $ \sgn(-\gamma x+{\rm O}(x^3))$          \\ 
 &\\
\hline

& \\  
Rhombs
       & $\sgn(b_1 + b_4), \quad \sgn(b_1 - b_4),\quad \sgn(\zeta_1),
		\quad \sgn(\zeta_2),$ \\ 
${\bf z}=(x,0,0,x,0,0)$ &
	where $\zeta_1+\zeta_2=(-2b_1-2b_4+2b_2+b_5+b_6)x^2$, \\ 
$0=\mu x + (b_1+b_4) x^3+{\rm O}(x^5)$
	& $\zeta_1\zeta_2=
	  -\gamma^2x^2+(b_1+b_4-b_2-b_5)(b_1+b_4-b_2-b_6)x^4$ \\ 
& \\ 

\hline
& \\ 
Rhombs
       & $\sgn(b_1 + b_5), \quad \sgn(b_1 - b_5),\quad \sgn(\zeta_1),
		\quad \sgn(\zeta_2),$ \\ 
${\bf z}=(x,0,0,0,x,0)$
	& where $\zeta_1+\zeta_2=(-2b_1-2b_5+2b_2+b_4+b_6)x^2$, \\  
$0=\mu x + (b_1+b_5) x^3+{\rm O}(x^5)$
	& $\zeta_1\zeta_2=
	  -\gamma^2x^2+(b_1+b_5-b_2-b_4)(b_1+b_5-b_2-b_6)x^4$ \\ 
& \\ 

\hline
& \\ 
Rhombs
       & $\sgn(b_1 + b_6), \quad \sgn(b_1 - b_6),\quad \sgn(\zeta_1),
		\quad \sgn(\zeta_2),$ \\ 
${\bf z}=(x,0,0,0,0,x)$
       & where $\zeta_1+\zeta_2=(-2b_1-2b_6+2b_2+b_4+b_5)x^2$ \\ 
$0=\mu x + (b_1+b_6) x^3+{\rm O}(x^5)$
       & $\zeta_1\zeta_2=-\gamma^2x^2
	 +(b_1+b_6-b_2-b_4)(b_1+b_6-b_2-b_5)x^4$ \\ 
& \\

\hline
& \\ & $\sgn(\gamma x+2(b_1+2b_2+b_4+b_5+b_6)x^2)$ \\

Super Hexagons
	& $\sgn(\gamma x+2(b_1+2b_2-b_4-b_5-b_6)x^2)$ \\ 
${\bf z}=(x,x,x,x,x,x)$ 
	& $\sgn(-\gamma x+{\rm O}(x^3)),
	  \quad \sgn(\zeta_1),\quad \sgn(\zeta_2)$,\\
$0=\mu x+\gamma x^2+(b_1+2b_2)x^3\quad$ 
	& where $\zeta_1+\zeta_2=-4\gamma x +4(b_1-b_2)x^2$ \\ 
$\qquad+(b_4+b_5+b_6)x^3+{\rm O}(x^4)$
	& $\zeta_1\zeta_2=4(\gamma x-(b_1-b_2)x^2)^2\qquad\qquad\qquad\quad$
\\
&$\qquad \qquad\qquad-2((b_4-b_5)^2+(b_4-b_6)^2+(b_5-b_6)^2))x^4$\\
&$\sgn(\zeta_0)$, where 
$\zeta_0={\rm O}(x^{2(m-1)})$\\ & \\
\hline
& \\
Super Triangles & \\
${\bf z} = (z,z,z,z,z,z),$ & Same as super hexagons \\
 $\quad z=x e^{i\psi}, \psi\ne 0,\pi,\ldots$ & except $\zeta_0 \to -\zeta_0$\\
$0=\mu z+\gamma \bar z^2+(b_1+2b_2)|z|^2z\quad$ & \\
$\qquad + (b_4 + b_5 + b_6)|z|^2z + {\rm O}(x^4)$ & \\
& \\

\hline
\end{tabular}
\end{table}

\bigskip

\section{Reduction to Bifurcation Equations}\label{sec:reduc}
\medskip

Although symmetry determines the form of the bifurcation equations, 
the particular
system determines $a_n$ (and $b_n$) 
in equations~(\ref{eq:sqeq}) (and~(\ref{eq:hexeq})).
Each coefficient $a_n$
is associated with a particular amplitude $z_n$ in~(\ref{eq:sqeq}), and
each of these amplitudes is in turn associated with a Fourier wave
vector ${\bf K}_n$.  In this section we use perturbation theory to
derive the coefficients
$a_n$, $\gamma$, and $b_n$ from the full system~(\ref{eq:full}).

The coefficients $a_1,\ldots,a_4$ and $b_1 (=a_1), b_4,b_5,b_6$ may all be
calculated by considering a single bifurcation problem involving just
two critical Fourier modes; the remaining coefficients $\gamma$ and
 $b_2$ are calculated by considering a bifurcation to hexagons.
To understand
the need for only two modes in the computation of $a_1,\ldots,a_4,
b_4,b_5,b_6$ consider, for example, the square lattice, where
the ${\bf K}_2, {\bf K}_3,$ and ${\bf K}_4$ vectors
are all related to the ${\bf K}_1$ vector
by an angle $\theta$.  Let $\theta_j$ denote the angle between ${\bf K}_1$
and ${\bf K}_j$; then
 $\theta_2=\pi/2$
and $\theta_4=\pi/2 + \theta_3$ 
where $\theta_3$ is related to the integer pair $(m,n)$ by
\eqno
\label{eq:costheta}
\cos\theta_3 = {2mn\over m^2 + n^2}
\endeqno
(see Figure~\ref{fig:lattice}).  
The coefficients $a_j$ are
a function of the angle $\theta_j$:
$$a_j = h(\theta_n),\quad j=2,3,4.$$
Similarly, on the hexagonal lattice, 
$$b_j = h(\theta_j), \quad j=4,5,6$$
where $\theta_5 = \theta_4 + 2\pi /3$ and $\theta_6 = \theta_4 + 4\pi /3$,
with $\theta_4$ determined by the integers $(m,n)$.  From symmetry
considerations we know that $h(\theta) = h(\pi - \theta)$.
A calculation with two critical modes will determine
$h(\theta)$, and hence determine all of the angle-dependent
coefficients in the bifurcation equations.

To compute $h(\theta)$ we seek solutions which are periodic on a rhombic
(or square) lattice; specifically, solutions of the form
\eqno
w_{\rm rh} = z_1 e^{i\k_1\cdot\r } + z_{\theta} e^{i \k_\theta\cdot\r}
	+ \cc,  \label{eq:cmrhomb}
\endeqno
where $\k_1 = q_c(1,0)$, 
$\k_\theta= q_c(\cos\theta, \sin\theta)$, and
$\theta$ is not a multiple of $\pi/3$.
The amplitudes evolve in a neighborhood of the bifurcation point
$\mu=0$ according to
\eqno
\dot z_1 &= &\mu z_1 + z_1(a_1 |z_1|^2 + h(\theta) |z_\theta|^2)\nonumber \\
\dot z_\theta &= &\mu z_\theta + z_\theta (a_1 |z_\theta|^2 
	+ h(\theta) |z_1|^2).
\label{eq:ztheta}
\endeqno
Stripes are solutions satisfying $z_\theta = 0$; rhombs correspond to
$z_1 = z_\theta$; squares correspond to $z_1 = z_{\pi/2}$.
Comparison of~(\ref{eq:ztheta}) with~(\ref{eq:sqeq})
restricted to the subspace ${\bf z}=(z_1,0,z_\theta,0)$ 
confirms the relationship $a_3=h(\theta_3)$, {\it etc.}

To compute $\gamma$ and $b_2$ in equation~(\ref{eq:hexeq}) we consider
a three-mode solution
\eqno
w_{\rm hex} = z_1 e^{i\k_1\cdot\r} + z_2 e^{i\k_2\cdot\r}
	+ z_3 e^{i\k_3\cdot\r} + \cc, \label{eq:cmhex}
\endeqno
where $\k_1 = q_c(1,0)$, $\k_2 = q_c(-1/2, \sqrt{3}/2)$, and
$\k_3 = q_c (-1/2, -\sqrt{3}/2)$.
The amplitudes evolve according to equations of the form
\eqno
&\dot z_1 = \mu z_1 + \gamma z_2^*z_3^* + z_1(b_1|z_1|^2 + 
	b_2(|z_2|^2 + |z_3|^2))&
\label{eq:zhex}
\endeqno
with similar equations for $\dot z_2$ and $\dot z_3$.

To compute the bifurcation equations~(\ref{eq:ztheta}) 
and~(\ref{eq:zhex}) from equation~(\ref{eq:full}) we first
Taylor-expand $f(u,v)$ and $g(u,v)$ through cubic order in $u$ and $v$:
\eqalign
	f(u,v) &= &au + bv + F_2(u,v) + F_3(u,v) + \cdots\\
	g(u,v) &= &cu + dv + G_2(u,v) + G_3(u,v) + \cdots .\\
\endalign
Here $F_2(u,v)$, $G_2(u,v)$, $F_3(u,v)$, and $G_3(u,v)$
denote the quadratic and cubic nonlinear terms:
\eqalign
  \left(\matrix{F_2(u,v)\cr G_2(u,v)\cr}\right) &\equiv
        &\left(\matrix{f_{uu} u^2/2 + f_{uv}u v + f_{vv}v^2/2\cr
                g_{uu} u^2/2 + g_{uv} u v + g_{vv} v^2/2\cr}\right),\\
\medskip
  \left(\matrix{F_3(u,v)\cr G_3(u,v)\cr}\right) &\equiv
        &\left(\matrix{f_{uuu}u^3/6 + f_{uuv}u^2 v/2 +
                f_{uvv} u v^2/2 + f_{vvv} v^3/6\cr
                g_{uuu}u^3/6 + g_{uuv}u^2 v/2 + g_{uvv}u v^2/2
                + g_{vvv} v^3/6\cr}\right),
\endalign
where the subscripts on $f$ and $g$ denote partial derivatives
evaluated at $u=v=0$.

We use perturbation theory to compute the small amplitude, slow evolution
on the center manifold.
Let
$$
   t_1 = \epsilon t, \quad t_2 = \epsilon^2 t,
$$
\eqno
&	\left(\matrix{u\cr v\cr}\right) = \epsilon 
	\left(\matrix{u_1\cr v_1\cr}\right)w_1(\r,t_1,t_2) + 
	\epsilon^2 \W_2 + \epsilon^3 \W_3 + O(\epsilon^4),&
\label{eq:w2eqn}
\endeqno
where $w_1$ takes the form of either~(\ref{eq:cmrhomb}) or~(\ref{eq:cmhex})
for the respective rhombic or hexagonal calculations,
%
%
and $\W_2$ and $\W_3$ represent higher order terms to be determined.
The vector $(u_1, v_1)^T$ is a null-vector of $L(q_c)$, at the bifurcation
point; see equation~(\ref{eq:lq}).
We further assume that the
linear parameters $a,b,c,d,K$ are within $O(\epsilon^2)$ of the
critical bifurcation values, {\it e.g.}
$a=a_c + \epsilon^2 \lambda_a$, {\it etc.} where $a_c, b_c, c_c, d_c, K_c$
satisfy~(\ref{eq:critparms}) and determine $q_c$ via substitution 
into~(\ref{eq:qc}).
With these assumptions, equation~(\ref{eq:full}) may be written as
\eqno
  \epsilon^2 {\partial\over \partial t_2} \left(\matrix{u\cr v\cr}\right) +
  \epsilon {\partial\over \partial t_1} \left(\matrix{u\cr v\cr}\right) +
	L_0\left(\matrix{u\cr v\cr}\right) =
	\epsilon^2 L_2 \left(\matrix{u\cr v\cr}\right) +
	\left(\matrix{F_2(u,v)\cr G_2(u,v)\cr}\right) +
	\left(\matrix{F_3(u,v)\cr G_3(u,v)\cr}\right)
	\label{eq:epseq}
\endeqno
plus higher order terms,
where
$$
  L_0 \equiv \left(\matrix{-\nabla^2 - a_c& -b_c\cr
			-c_c &-K_c\nabla^2 - d_c\cr}\right)
$$ 
and
$$
  L_2 \equiv
	\left(\matrix{ \lambda_a &\lambda_b\cr 
	\lambda_c &\lambda_d + \lambda_K\nabla^2\cr}\right).
$$


\subsection{Rhombic Lattice Computation}
For the rhombic case,
the equation for $\W_2$ is given at $O(\epsilon^2)$ as
\eqno
  \left(\matrix{u_1\cr v_1\cr}\right) 
     {\partial \over \partial t_1} w_{\rm rh}+
	L_0 \W_2 =
	\left(\matrix{F_2(u_1,v_1)\cr G_2(u_1,v_1)\cr}\right)
	w_{\rm rh}^2,\label{eq:rheps2}
\endeqno
where $w_{\rm rh}$ is given by~(\ref{eq:cmrhomb}).
Through the term $w_{\rm rh}^2$ the quadratic nonlinearities
generate four terms with distinct wave numbers:
\eqalign
	w_2 &= &z_1^2 e^{i2\k_1\cdot\r} + z_\theta^2 e^{i2\k_\theta\cdot\r}+\cc,
		\quad \nabla^2 w_2 = -4q_c^2w_2, \\
	w_3 &= &z_1z_\theta e^{i (\k_1 + \k_\theta) \cdot \r} + \cc,
		\quad \nabla^2 w_3 = - 2q_c^2(1 + \cos\theta)w_3, \\
	w_4 &= &z_1z_\theta^* e^{i (\k_1 - \k_\theta)\cdot\r} + \cc,
		\quad \nabla^2 w_4 = -2q_c^2(1 - \cos\theta)w_4,\\
	w_5 &= &|z_1|^2 + |z_\theta|^2, \quad\nabla^2 w_5 = 0.\\
\endalign
There are no terms on the right hand side of~(\ref{eq:rheps2})
with wave number $q_c$, so the solvability condition is
$${\partial\over \partial t_1} w_{\rm rh} = 0.$$
Since there is no dependence on the
$t_1$-time scale it will be omitted from the remaining analysis.
Solving equation~(\ref{eq:rheps2}) determines
\eqno
  \W_2 \equiv \left(\matrix{U_2\cr V_2\cr}\right) =
  \sum_{n=2}^5 \left(\matrix{u_n\cr v_n\cr}\right) w_n,
  \label{eq:w2def}
\endeqno
where $u_n$ and $v_n$ ($n=2,\ldots, 5$) are constant and satisfy
\eqno
  L_0\left(\matrix{u_n\cr v_n\cr}\right)w_n = 
	\left(\matrix{F_2(u_1,v_1)\cr G_2(u_1,v_1)}\right) c_n w_n,
	\label{eq:l0w2}
\endeqno
where $c_2=1$ and $c_3=c_4=c_5=2$. 
Explicit expressions for $u_n$
and $v_n$ are given in the Appendix.

The equation at $O(\epsilon^3)$ is
\eqno
  \left(\matrix{u_1\cr v_1\cr}\right) 
  {\partial \over \partial t_2} w_{\rm rh}+
	L_0 \W_3 & = &
	L_2 \left(\matrix{u_1\cr v_1\cr}\right) w_{\rm rh} +
	\left(\matrix{{\partial F_2(u_1,v_1)\over\partial u_1}\cr 
		      {\partial G_2(u_1,v_1)\over\partial u_1}\cr}\right)
		U_2 w_{\rm rh} \nonumber \\
&&	+ \left(\matrix{{\partial F_2(u_1,v_1)\over\partial v_1}\cr 
		      {\partial G_2(u_1,v_1)\over\partial v_1}\cr}\right)
		V_2 w_{\rm rh} \nonumber \\
&&	+ \left(\matrix{F_3(u_1,v_1)\cr G_3(u_1,v_1)\cr}\right) w_{\rm rh}^3,
\label{eq:rheps3}
\endeqno
where $(U_2, V_2)$ are given by~(\ref{eq:w2def}).
There are now terms on the right hand side of
the equation with spatial dependence $e^{i\k_1\cdot\r}$ 
and $e^{i\k_\theta\cdot\r}$; the operator $L_0$ is not invertible
for terms with this Fourier dependence.  To ensure that these terms will lie
in the range of $L_0$ we apply the Fredholm alternative theorem
to obtain the solvability condition.  In this computation we use the
fact that the nullspace of the adjoint linear operator $L_0^\dagger$
is spanned by
$$(\tilde u, \tilde v) e^{i\k_1\cdot\r},
	\quad (\tilde u,\tilde v) e^{i\k_\theta\cdot\r},
$$
 where
\eqno
  &\left(\tilde u,\tilde v\right)=\left(c, -a_c+q_c^2\right).&
	\label{eq:nullvec}
\endeqno
The inner product of equation~(\ref{eq:rheps3}) with these left null-vectors
yields the solvability condition, which
takes the expected form of the bifurcation
equations~(\ref{eq:ztheta}).
After rescaling time to absorb the factor
$u_1\tilde u + v_1\tilde v$, which is always positive,
we obtain the rhombic bifurcation equation coefficients
\eqno
\label{eq:a_1}
a_1= (u_2+u_5)\eta_1 + (v_2+v_5)\eta_2 + 3\beta
\endeqno
and
\eqno
\label{eq:htheta}
h(\theta) = (u_3+u_4+u_5)\eta_1 + (v_3+v_4+v_5)\eta_2 + 6\beta,
\endeqno
where
\eqno
\eta_1 &\equiv &\tilde u {\partial F_2(u_1,v_1)\over \partial u_1}
        + \tilde v {\partial G_2(u_1, v_1)\over \partial u_1},\nonumber\\ 
\eta_2 &\equiv &\tilde u {\partial F_2(u_1,v_1)\over \partial v_1}
        + \tilde v {\partial G_2(u_1, v_1)\over \partial v_1},
	\label{eq:etabeta}\\
\beta &\equiv &\tilde u F_3(u_1,v_1) + \tilde v G_3(u_1, v_1).\nonumber
\endeqno
The expressions for $u_n$ and $v_n$, $n=2,\ldots, 5$, are given in the
Appendix.



\subsection{Hexagonal Lattice Computation}
	The hexagonal calculation proceeds in a fashion similar to the
rhombic case, but requires some additional work due to the presence of 
resonant terms with wave number $q_c$
which are generated by the quadratic nonlinearities.

To simplify the presentation we let $z_1=z_2=z_3=z_h$ in~(\ref{eq:cmhex}).
Since 
the form of the amplitude equation is known from equation~(\ref{eq:hexeq}),
and $b_1 (=a_1)$ is known from the rhombic lattice calculation,
the remaining coefficients $\gamma$ and $b_2$
may be extracted from the quadratic and cubic coefficient expressions,
respectively.

At $O(\epsilon^2)$ the equation to be solved is
\eqno
  \left(\matrix{u_1\cr v_1\cr}\right) {\partial \over \partial t_1}\whex +
	L_0 \W_2 &=
	&\left(\matrix{F_2(u_1,v_1)\cr G_2(u_1,v_1)\cr}\right)
	\whex^2\label{eq:hexeps2}
\endeqno
where $\whex$ is given by~(\ref{eq:cmhex}).
The term $\whex^2$ generates four terms 
with distinct wave numbers:
\eqalign
	w_2 &= &z_h^2 (e^{i2\k_1\cdot\r} + e^{i2\k_2\cdot\r}
		+ e^{i2\k_3\cdot\r}) +\cc,\quad \nabla^2 w_2 = -4q_c^2w_2,\\
	w_5 &= &|z_h|^2, \quad\nabla^2 w_5 = 0,\\
	w_6 &= &|z_h|^2 (e^{i(\k_1-\k_2)\cdot\r} + 
		e^{i(\k_2-\k_3)\cdot r} + e^{i(\k_3-\k_1)\cdot\r}) + \cc,
		\quad \nabla^2 w_6 = - 3q_c^2w_6,\\
	w_7 &= &{z_h^*}^2 (e^{i \k_1\cdot\r} + e^{i\k_2\cdot\r}
		+ e^{i\k_3\cdot\r}) + \cc, \quad \nabla^2 w_7 = -q_c^2 w_7.
\endalign
From $w_7$ we see that there are now terms on the right hand side
of~(\ref{eq:hexeps2}) 
with wave number $q_c$, for which $L_0$ is
not invertible.
Unlike in the rhombic case the $t_1$ time scale is necessary here,
reflecting that the small-amplitude
dynamics are dominated by the quadratic nonlinearities.
As in the rhombic calculation
the Fredholm alternative theorem is
applied, giving the solvability condition
\eqno
(u_1\tilde u + v_1 \tilde v) {\partial \over \partial t_1} z_h&= 
	&\gamma {z_h^*}^2,
	\qquad \gamma = 2(\tilde u F_2(u_1, v_1) + \tilde v G_2(u_1, v_1)).
\label{eq:quadsolve}
\endeqno
The factor $(u_1\tilde u + v_1\tilde v)$ is
positive (from conditions~(\ref{eq:critparms}) and~(\ref{eq:negcond}))
and will later be absorbed by a rescaling of time.
Note that if $\gamma$ is $O(\epsilon)$ then the $t_1$-time scale is
unnecessary.  This is the situation of interest here, since we know
from Table~\ref{tab:evhex} that when $\gamma$ is $O(1)$ there are no
stable small-amplitude steady states.  However, in the following computation
we must retain both time scales to compute the general form of the
cubic coefficient $b_2$.  In our subsequent analysis we will focus
on the degenerate problem $\gamma = 0$.

Continuing as in the rhombic case,
the solution of~(\ref{eq:hexeps2}) is
$$
  \W_2 \equiv \left(\matrix{U_2\cr V_2}\right) =
   \sum_{n=2,5,6,7} \left(\matrix{u_n\cr v_n\cr}\right) w_n,
$$
where $u_2$, $v_2$, $u_5$, and $v_5$ are
known from the rhombic lattice computation, and
$u_6$ and $v_6$ are computed in a similar fashion.
Equations~(\ref{eq:hexeps2}) and~(\ref{eq:quadsolve}) 
lead to the following solvable equation for 
$(u_7,v_7)$:
$$L_0 \left(\matrix{u_7\cr v_7\cr}\right) w_7 = 
	-{\gamma\over u_1\tilde u + v_1\tilde v} 
	  \left(\matrix{u_1\cr v_1\cr}\right) w_7 +
	2\left(\matrix{F_2(u_1, v_1)\cr G_2(u_1, v_1)\cr}\right) w_7.
$$
%
We choose the particular solution of this singular problem to be
$$u_7=0,\quad v_7 = -{1\over b_c}\left(-{\gamma u_1\over 
	u_1\tilde u + v_1\tilde v} + 2F_2(u_1,v_1)\right).
$$

	At $O(\epsilon^3)$ the equation is
\eqno
\left(\matrix{u_1\cr v_1\cr}\right) {\partial\over\partial t_2} \whex + 
	{\partial\over \partial t_1}\W_2 +
	L_0 \W_3 &=
	&L_2 \left(\matrix{u_1\cr v_1\cr}\right) \whex \cr
&&	+ \left(\matrix{{\partial F_2(u_1,v_1)\over\partial u_1}\cr 
		      {\partial G_2(u_1,v_1)\over\partial u_1}\cr}\right)
		U_2 \whex \nonumber \\
&&	+ \left(\matrix{{\partial F_2(u_1,v_1)\over\partial v_1}\cr 
		      {\partial G_2(u_1,v_1)\over\partial v_1}\cr}\right)
		V_2 \whex \cr
&&	+ \left(\matrix{F_3(u_1,v_1)\cr G_3(u_1,v_1)\cr}\right) \whex^3.
\label{eq:hexeps3}
\endeqno
Again the Fredholm alternative theorem must be applied, giving a
solvability condition of the form
$$
  (u_1\tilde u + v_1\tilde v){\partial\over\partial t_2}z_h = 
	\lambda z_h + (b_1 + 2b_2)|z_h|^2z_h.
$$
%
%
By letting $z(t) = \epsilon z_h(t_1, t_2)$, $\mu = \epsilon^2 \lambda$,
and rescaling time to absorb the factor $u_1\tilde u + v_1\tilde v$
on the left hand side,
we obtain the reconstituted bifurcation equation
$$\dot z = \mu z + \gamma {z^*}^2 + (b_1 + 2b_2) |z|^2 z.$$
The coefficients $\mu$, $\gamma$, $b_1$, and $b_2$ are given in the Appendix.

\section{Parameter Collapse}\label{sec:collapse}

%



	Naively one might expect that the cubic coefficients $a_n$
and $b_n$ in the bifurcation equations
may be adjusted to take on
any desired value, by varying the many nonlinear coefficients
in the Taylor expansions of $f(u,v)$ and $g(u,v)$.
However, the expressions for $a_n$ and $b_n$,
given in Appendix~\ref{ap:coeffs},
clearly show that all eight cubic 
coefficients contained within $F_3(u_1,v_1)$ and
$G_3(u_1,v_1)$ collapse into the {\em single} expression
$\beta \equiv \tilde u F_3(u_1,v_1) + \tilde v G_3(u_1, v_1)$.
Similarly, the six quadratic coefficients in $F_2(u,v)$ and $G_2(u,v)$
are contained in the coefficient expressions solely in the 
four terms $F_2(u_1,v_1)$, $G_2(u_1, v_1)$, $\eta_1$ and $\eta_2$, 
given by~(\ref{eq:etabeta}).
Moreover, these last four quantities satisfy the relation
\eqno
  &u_1 \eta_1 + v_1 \eta_2 = 2 \tilde u F_2(u_1,v_1) + 2\tilde v G_2(u_1,v_1).&
  \label{eq:degen1}
\endeqno
Therefore the six quadratic coefficients collapse into {\em three} independent 
parameters.
	In summary, the fourteen nonlinear coefficients in the Taylor
expansions of $f(u,v)$ and $g(u,v)$ enter our analysis in just four
independent combinations.

\subsection{Hexagonal degeneracy}\label{sec:hexdegen}

The parameter collapse is even more dramatic at the degeneracy $\gamma=0$ in
the hexagonal bifurcation problem~(\ref{eq:hexeq}).
An additional striking phenomena occurs at this point, which 
results in all $\theta$-dependence of the coefficients dropping out,
thereby severely restricting the types of patterns which may bifurcate
stably. Below we show that the bifurcation equations depend on just
two quantities determined from the reaction kinetics
of~(\ref{eq:full}).  Thus, we may simply and completely characterize
the possible bifurcation scenarios for~{\em all} reaction-diffusion
systems of the form~(\ref{eq:full}) restricted to the hexagonal
bifurcation problem at the degeneracy.

The degeneracy condition is
\eqno
  \gamma \equiv \tilde u F_2(u_1,v_1) + \tilde v G_2(u_1,v_1) = 0,
  \label{eq:hexdegen}
\endeqno
which implies
\eqno
  u_1 \eta_1 + v_1 \eta_2 = 0\label{eq:blah1}
\endeqno
by substitution into~(\ref{eq:degen1}).  
These two equations may be
solved for $F_2$ and $\eta_1$ in terms of $G_2$ and $\eta_2$, and
therefore
these four parameters collapse into the single parameter combination
$\eta_2 G_2(u_1,v_1)$ in 
the coefficient expressions for $a_n$ and $b_n$.

\begin{figure}
  \epsfbox{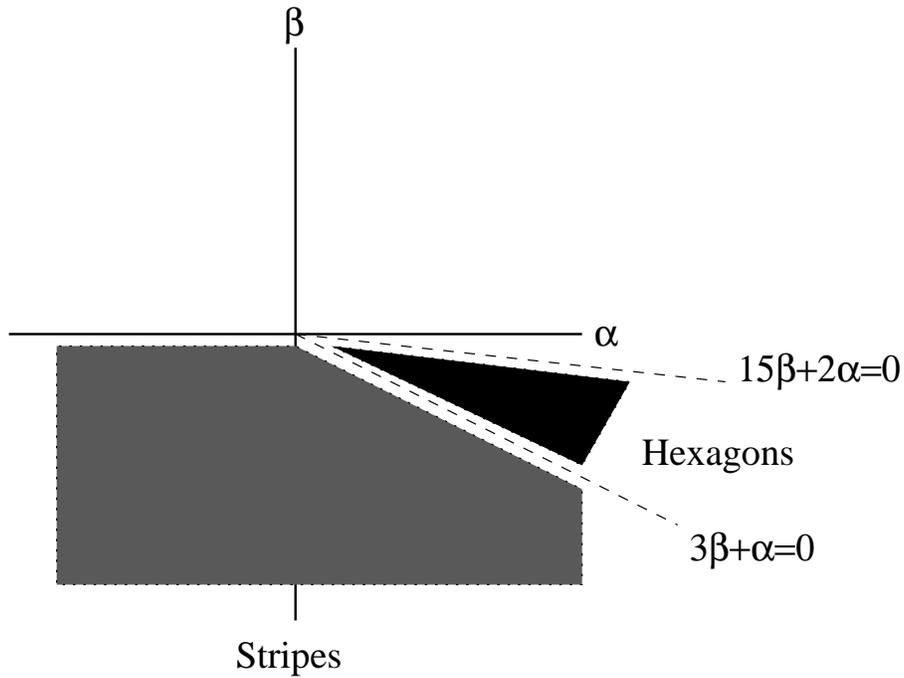}
  \caption{Stability of patterns at the hexagonal degeneracy $\gamma=0$.
	The details of 
	$f(u,v)$ and $g(u,v)$ in~(\ref{eq:full})
	enter the stability analysis via
	two effective parameters: $\alpha$, which depends on the quadratic
	nonlinearities, and $\beta$, which depends on the cubic nonlinearities.
	Rhombs and super hexagons are always unstable, as shown in
	Section~\protect\ref{sec:hexdegen}.  All patterns bifurcate unstably 
	for $\beta>0$ and/or $15\beta + 2\alpha > 0$.
	For the region labeled ``hexagons'' we have determined only
that hexagons are neutrally stable, since their stability  for
$\gamma=0$
depends on terms higher than cubic in~(\protect\ref{eq:hexeq}).
  }
  \label{fig:alphabet}
\end{figure}

The bifurcation coefficients $b_1=a_1$, $h(\theta)$, and $b_2$ contain
the terms $u_2, v_2, \ldots, u_6, v_6$, as given in the Appendix
(see also equations~(\ref{eq:a_1}) and~(\ref{eq:htheta})).
These terms are determined
by solving equations 
of the form ({\it cf.}~(\ref{eq:l0w2}))
$$-L_0(q_c)\left(\matrix{u_n\cr v_n\cr}\right)
	+ q_c^2f_n(\theta)\left(\matrix{1 &0\cr 0&K\cr}\right)
	  \left(\matrix{u_n\cr v_n\cr}\right)
	= c_n\left(\matrix{F_2(u_1,v_1)\cr G_2(u_1,v_1)\cr}\right).$$
Here $L_0(q_c)$ is the singular matrix~(\ref{eq:lq}) 
from the linear calculation, $c_n$ is a constant, and $f_n(\theta)$
is nonzero; for example, $f_3(\theta) = 1+2\cos\theta$.
Multiplying the above equation by the left null-vector
$(\tilde u,\tilde v)$ of $L_0(q_c)$ gives
$$\tilde u u_n + K\tilde v v_n = 0$$
at the degeneracy.  Using this relation and relation~(\ref{eq:blah1}),
we see that
$$u_n \eta_1 + v_n\eta_2 = 
	v_n\eta_2\left({K v_1 \tilde v\over u_1 \tilde u} + 1\right).
$$
From the definitions of the null-vectors $(u_1, v_1)$ and 
$(\tilde u,\tilde v)$
it immediately follows that
$$u_n\eta_1 + v_n\eta_2 = 0,\quad n=2,\ldots,6.$$
These expressions may then be substituted into the bifurcation
coefficient expressions from the Appendix, giving
$$  b_1\equiv (u_2+u_5)\eta_1 + (v_2+v_5)\eta_2 + 3\beta = 3\beta,$$
$$
  h(\theta) \equiv (u_3+u_4+u_5)\eta_1 + (v_3+v_4+v_5)\eta_2 + 6\beta = 6\beta,
$$
$$  b_2\equiv (u_5 + u_7)\eta_1 + (v_5+v_6+v_7)\eta_2 - 2\gamma (\tilde v v_6)
        + 6\beta = v_7\eta_2 + 6\beta,$$
for $\gamma=0$.  In short,
{\em the quadratic nonlinearities from $F_2(u,v)$ and $G_2(u,v)$ completely 
disappear} from the
bifurcation coefficients $b_1$ and $h(\theta)$, and appear in $b_2$ solely
through the term
\eqno
  \alpha \equiv v_7\eta_2. \label{eq:alpha}
\endeqno
This further
implies that all $\theta$-dependence disappears from the
bifurcation equations, and that rhombs and super hexagons are
{\em always} unstable at the bifurcation point, for the degenerate
problem.  
The signs of the eigenvalues for rhombs and super hexagons are given in 
Table~\ref{tab:evhex}.  The first two eigenvalues for
rhombs depend on the signs of
$b_1+b_4$ and $b_1-b_4$, which cannot both be negative.  Hence rhombs
are always unstable.  The signs of the first three super hexagon eigenvalues
are determined by the signs of
$$e_1 \equiv b_1 + 2b_2 + b_4 + b_5 + b_6 = 33\beta + 2\alpha,$$
$$e_2 \equiv b_1 + 2b_2 - b_4 - b_5 - b_6 = -3\beta + 2\alpha,$$
$$e_3 \equiv b_1-b_2 = -3\beta - \alpha.$$
All three cannot be made negative,
since $e_1 = -(3e_2+8e_3)$.
Stripes and hexagons may bifurcate stably;
Figure~\ref{fig:alphabet} gives 
their stability assignments in the $(\alpha,\beta)$-parameter plane.
Note that, for the cubic truncation of the bifurcation problem,
hexagons are only known to be neutrally stable, since
the eigenvalue $-\gamma x + O(x^3)$ in Table~\ref{tab:evhex} depends
on higher-order terms, which are not computed here.
For the corresponding square lattice computation, stripes are stable for
$\beta<0$; all other solutions are unstable for $\gamma=0$.  This picture
can be made even simpler for systems with $F_2(u,v) \propto G_2(u,v)$,
such as the CIMA model considered in the next section.  In these systems,
$\gamma=0$ typically implies $F_2(u_1,v_1) = G_2(u_1,v_1) = 0$, leading
to $\alpha=0$ in Figure~\ref{fig:alphabet}.
Stripes are then stable provided $\beta < 0$.

\begin{figure}
\epsfxsize=340pt
\epsffile{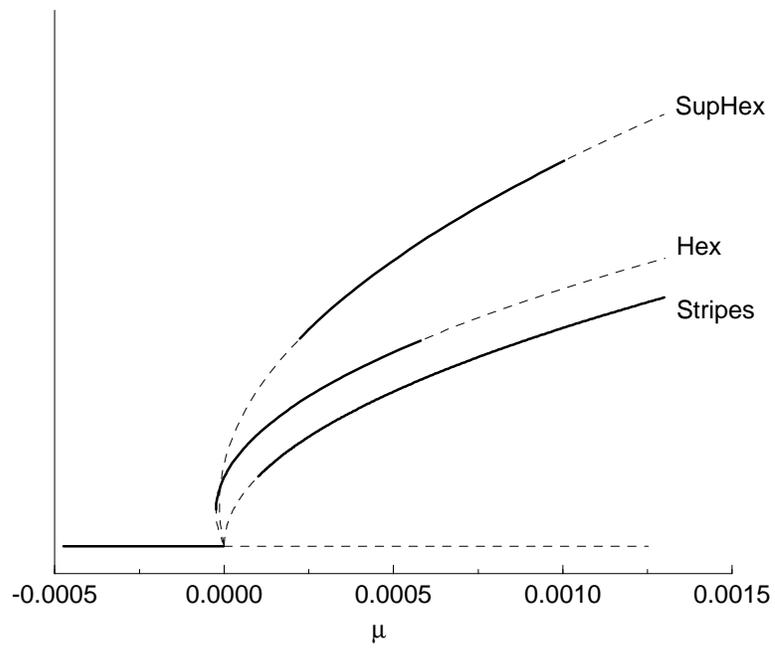}
\caption{Hexagonal lattice bifurcation diagram for the reaction-diffusion 
	system with $\gamma = 0.03$,
	$\alpha = -2$, and $\beta = -1/3$, as described in 
	Section~\ref{sec:hexdegen}.  Secondary branches and 
	unstable primary branches are not
	plotted.  Clearly there is a range of $\mu$ for which this
	system is tri-stable.
	}
\label{fig:abunfold}
\end{figure}

\begin{figure}
\epsfxsize=250pt\epsffile{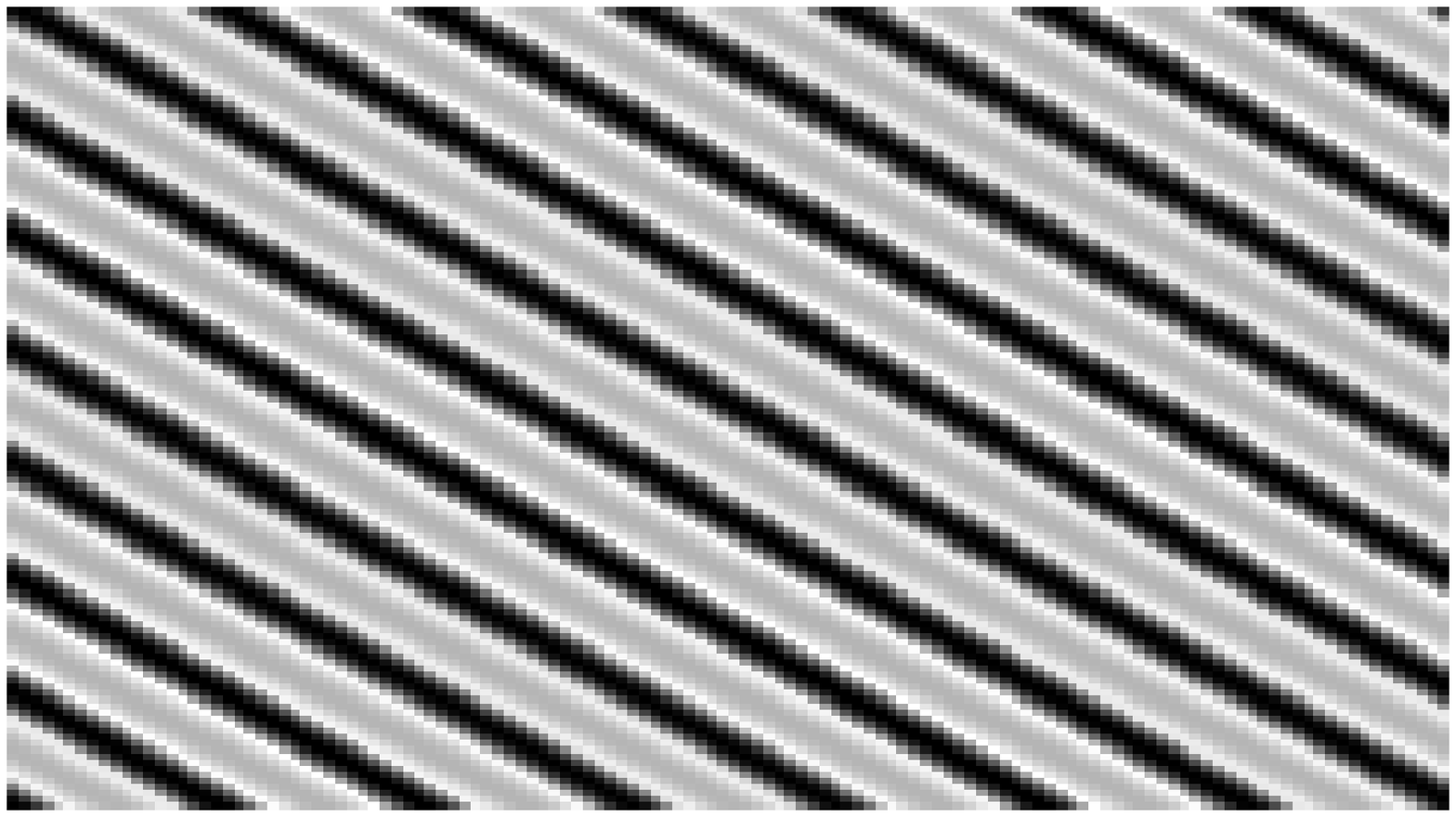}
\smallskip
\epsfxsize=250pt
\epsffile{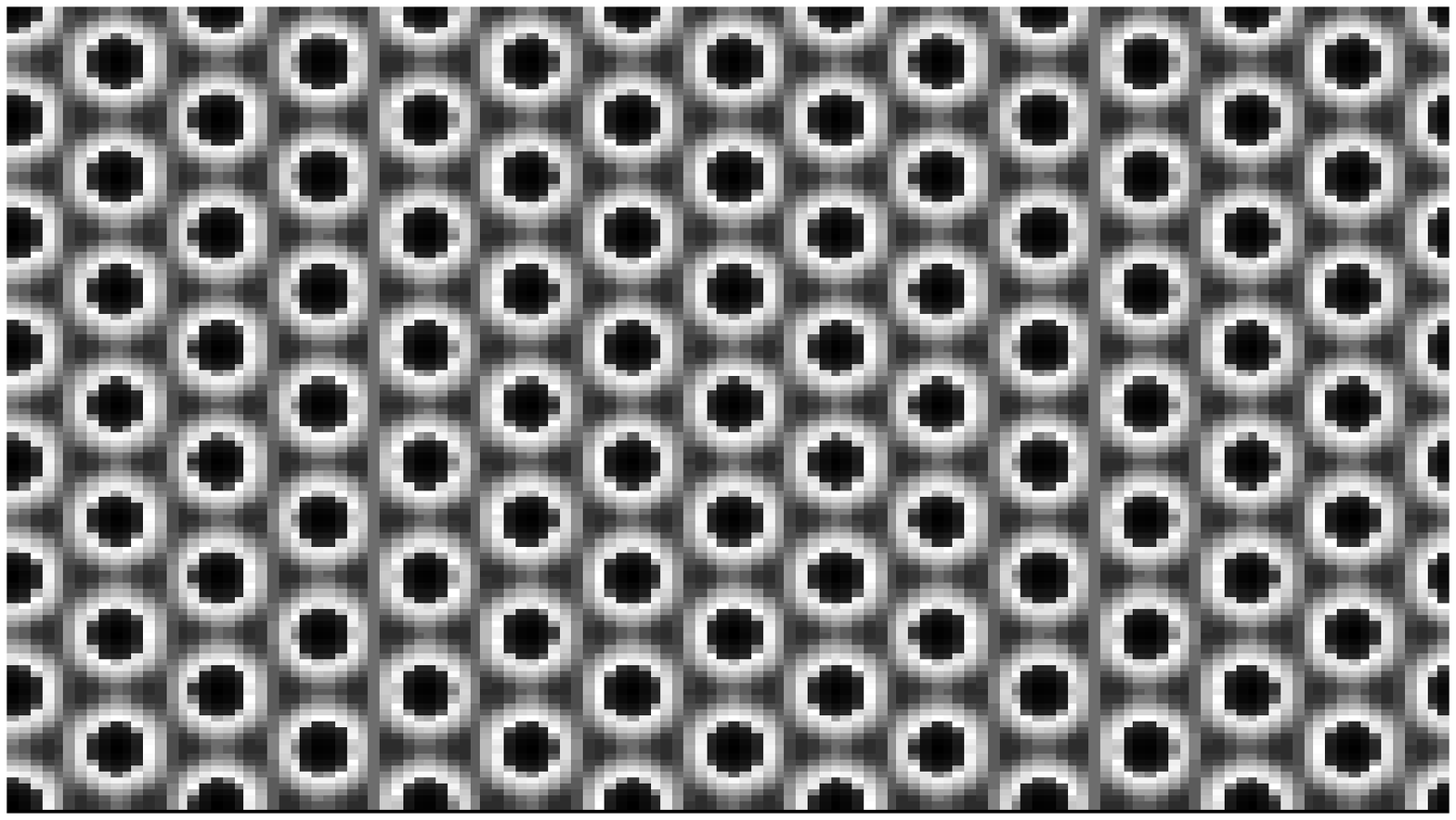}
\smallskip
\epsfxsize=250pt
\epsffile{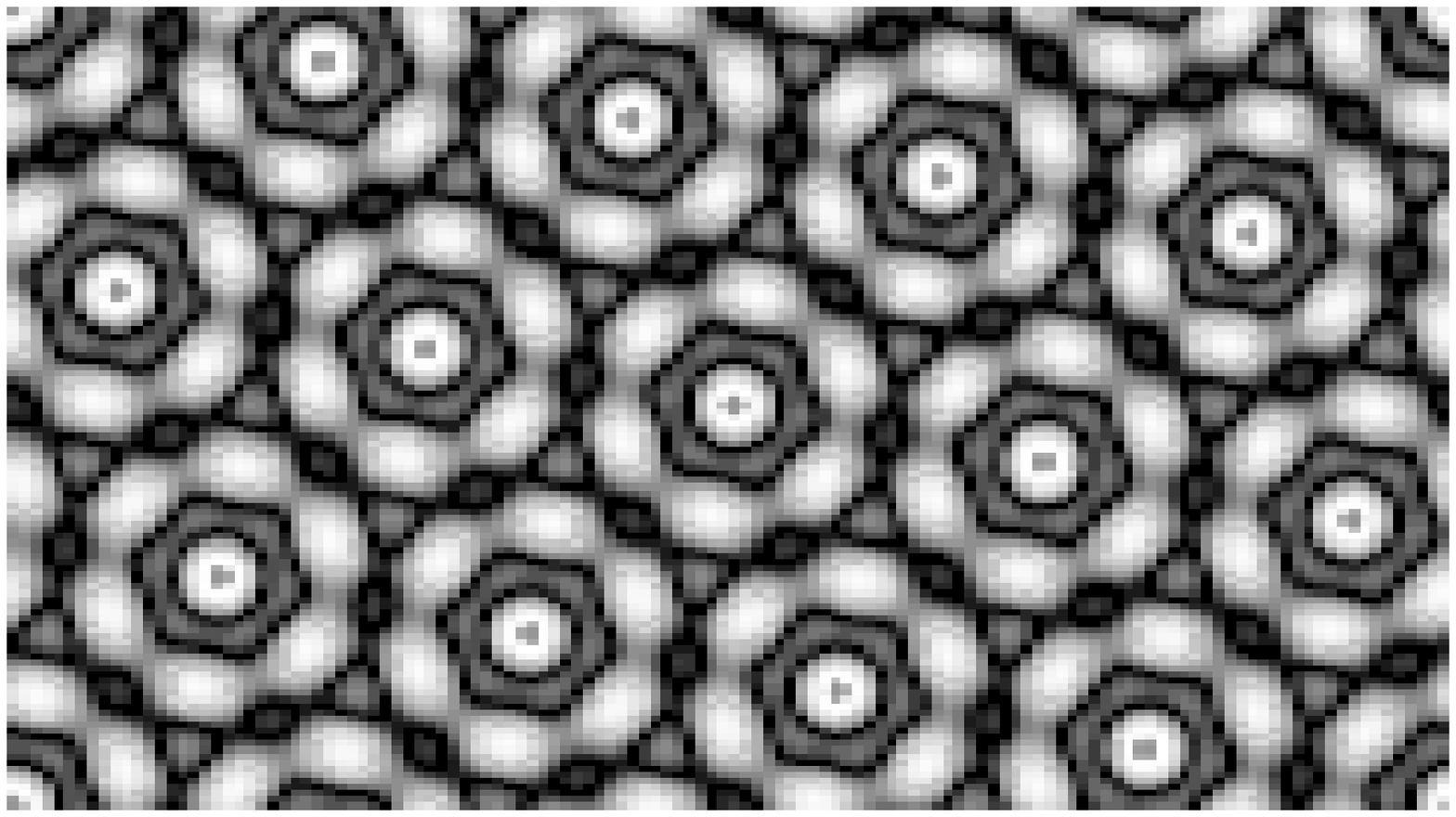}
\caption{The 
  steady state concentration field $u(\protect{\bf r})$ in the form
  of stripes, hexagons, and super hexagons, as computed numerically 
	from the reaction-diffusion system~(\protect\ref{eq:full}) with
  reaction-kinetics
	given by~(\protect\ref{eq:tristab}).  The sole difference between
	these numerical runs was in the initial condition.  Compare
	with the bifurcation diagram given in
  Figure~\protect\ref{fig:abunfold}.
  Numerical simulations were performed with a pseudo-spectral
  Crank-Nicholson
  scheme, using second-order Adams-Bashforth on the nonlinear terms.
  The rectangular domain has aspect ratio 
  $\protect\sqrt{3}$, with a 128$\times$128 grid.
}
\label{fig:tristab}
\end{figure}

We next unfold the degenerate problem by considering
$0< |\gamma| \ll 1$ in~(\ref{eq:cmhex}), keeping $b_1=3\beta$,
$b_2 = \alpha + 6\beta$, and $b_4 = b_5 = b_6 = 6\beta$.
There is now a small range of values $\mu \in (\mu_1, \mu_2)$ for which
super hexagons (or triangles) are stable, provided that $\alpha<0$,  
$\beta \in (\alpha/3, -\alpha/21)$.  For example, 
Figure~\ref{fig:abunfold} shows
the unfolded bifurcation diagram for $\beta=-1/3$, $\alpha=-2$
$(0 < \gamma \ll 1)$.
It is straightforward to construct a system of the form~(\ref{eq:full})
with these values.  Let
$$K=4,$$
\eqno
f(u,v) = (2+\lambda_a)u - 1.8v + 0.2378u^2 - 0.0141u^3,\label{eq:tristab}
\endeqno
$$g(u,v) = 4.05u - 2.8v - 0.04u^2 + uv.$$ This system gives
$\gamma\approx 0.03$, $\alpha\approx -2$, $\beta\approx -1/3$, and
according to the unfolding calculation gives stable super hexagons (or
triangles) in the range $3.07\times 10^{-5} < \lambda_a < 1.38\times
10^{-4}$.  A value of $\lambda_a = 7.6\times 10^{-5}$ corresponds to
$\mu = 5.54\times 10^{-4}$ in Figure~\ref{fig:abunfold}, at which
point rolls, hexagons, and super hexagons are all stable according to
our analysis.  Figure~\ref{fig:tristab} shows a numerical integration
of this system with $\lambda_a = 7.6\times 10^{-5}$, using a
pseudo-spectral Crank-Nicholson scheme on a $128\times 128$
rectangular grid of aspect ratio $\sqrt{3}$.  All three pictures are
at the same parameter values, and differ solely in the initial
condition.  The amplitude of the fundamental mode in each case is
within two percent of the steady-state values predicted by the
amplitude equations.

\bigskip

\section{Application to Model Equations}\label{sec:applications}

\subsection{Lengyel-Epstein model}\label{sec:cima}

	The results of the previous sections
will now be applied to a reaction-diffusion system
proposed by Lengyel and Epstein~\cite{ref:lengep1}
as a reduced model for the CIMA reaction:
\eqno
  \begin{array}{rcl}
	A_t &= &\hat a - A - {4AB\over 1+A^2} + \nabla^2 A\cr
	B_t &= &\delta \hat b\left(A - {AB\over 1+A^2}
			+ {\hat c\over\hat b}\nabla^2 B\right).
  \end{array}
& \label{eq:cimablah}
\endeqno
Note that the nonlinear terms differ solely by a multiplicative constant,
which will further simplify the hexagonal lattice computations.

Our analysis assumes a steady-state solution of $u=v=0$.  Thus we
let
$$
  A = x_0 + (1+x_0^2) u, \quad B = (1+x_0^2)(1+v)$$
where
$$x_0 \equiv {\hat a\over 5} > 0.$$
After expanding
the nonlinearities to cubic order the equation may be written as
\eqno
  \begin{array}{rcl}
	u_t &= &\nabla^2 u + au + bv + F_2(u,v) + F_3(u,v)\cr
	v_t &= &\delta \hat b[K\nabla^2 v + cu + dv + G_2(u,v)+ G_3(u,v)],
  \end{array}
& \label{eq:cima}
\endeqno
after rescaling time and space to absorb a factor of $1+x_0^2$.
The coefficients in~(\ref{eq:cima}) 
are related to those in~(\ref{eq:cimablah}) by
$$a=3x_0^2-5,\quad b=-4x_0,\quad c=2x_0^2,\quad d=-x_0,\quad K={\hat
c\over \hat b}$$
$$F_2(u,v) = 4x_0(3-x_0^2)u^2 + 
  4(x_0^2-1) uv,$$
$$F_3(u,v) = 4(x_0^4 - 6x_0^2 + 1) u^3 +
  4x_0 (3-x_0^2) u^2v,$$
$$G_2(u,v) = {1\over 4} F_2(u,v),\quad G_3(u,v) = {1\over 4} F_3(u,v).$$
The overall factor of $\delta \hat b$ in the $v_t$-equation
will later be removed from the bifurcation problem
by rescaling time.  That is, although the overall factor of $\delta 
\hat b$ enters the linear calculations -- in particular the 
necessary conditions~(\ref{eq:negcond}) for Turing pattern formation -- 
it simply scales out of the final bifurcation problem.
In addition to the two remaining equation parameters, $\hat a$ and $K$,
the lattice angle $\theta$ will enter our bifurcation analysis. 
Choosing $K$ to fix the system at the Turing bifurcation point,
the resulting bifurcation scenario is then determined by $\hat a$ and
$\theta$,
where, for example, $\theta$ is related to the choice of square
lattice
by~(\ref{eq:costheta}).
\bigskip

\begin {figure}
  \epsfxsize=360pt\epsffile{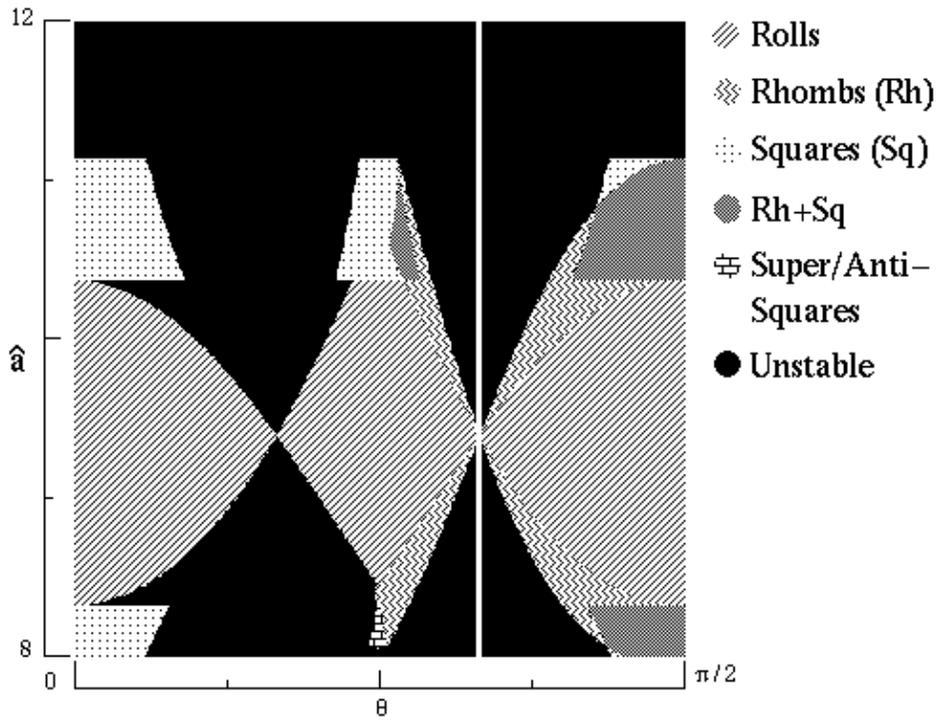} 
  \caption{Square lattice bifurcation
  results (for various lattice angles $\theta$) as a function of
  the parameter $\hat a$ in the Lengyel-Epstein CIMA 
	model~(\protect\ref{eq:cimablah}).  The diagram
  was generated by substituting the coefficient expressions from
  Appendix~\ref{ap:coeffs} into Table~\ref{tab:evsq} and evaluating at
  different points in the $\hat a - \theta$ plane.  The region near
  lattice angle $\theta=\pi/3$ has been removed,
  since this angle gives a hexagonal interaction,.
  Note also that at $\hat a\approx 9.4$ rolls are stable for all
  $\theta$. This value of $\hat a$ yields
  the degeneracy $\gamma = 0$ in
  the hexagonal bifurcation problem~(\ref{eq:hexeq}).}  
\label{fig:cimaparms}
\end{figure}

\begin{figure}[h]
\begin{center}
\begin{picture}(60,60)
        \put(30,30){\circle{40}}
        \put(30,30){\vector(2,1){18}}
        \put(30,30){\vector(-2,1){18}}
        \put(48,39){\vector(-2,1){18}}
        \put(30,48){\circle*{2}}
        \put(50,38){$k_1=(2,1)$}
        \put(2,38){$k_4$}
        \put(31,52){$k_1+k_4$}
        \put(0,30){\line(1,0){60}}
        \put(30,0){\line(0,1){60}}
\end{picture}
\end{center}
\caption{Critical modes $(2,1)$ and $(-2,1)$  on a
$(2,1)$-square lattice interact nonlinearly to generate a mode (0,2)
that lies very near to the critical circle. We expect that this proximity 
will greatly reduce the domain of validity of the bifurcation calculation.}
\label{fig:reson}
\end{figure}
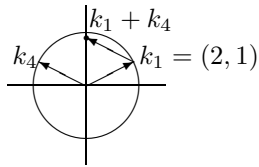

We first consider the relative stability of
patterns which lie on  a square lattice.  
Figure~\ref{fig:cimaparms} is a plot in the $(\theta,\hat a)$ plane,
generated by calculating the bifurcation coefficients $a_1$, $a_2=h(\pi/2)$,
$a_3 = h(\theta)$, and $a_4 = h(\pi/2 - \theta)$, as given in the Appendix.
These coefficients are then substituted into the eigenvalue 
expressions in Table~\ref{tab:evsq}, generating the stability assignments
in the figure.  The region surrounding $\theta = \pi/3$ has been removed
since $h(\theta)$ diverges as $\theta\to \pi/3$; the hexagonal lattice 
analysis is required at this point.  Also, in the vicinity of
$\theta = \pi/3$ the
domain of validity of the bifurcation results is very small;
certain slaved modes are only weakly damped, 
leading to a small-divisor problem in the computation of 
$h(\theta)$ (see Figure~\ref{fig:reson} for an example).
Finally, we emphasize that anti- and super square patterns only exist for a
discrete set of $\theta$ values, as discussed in Section~\ref{sec:formulation}.

\bigskip




\begin {figure}
\medskip
  \epsfxsize = 6in \epsfbox{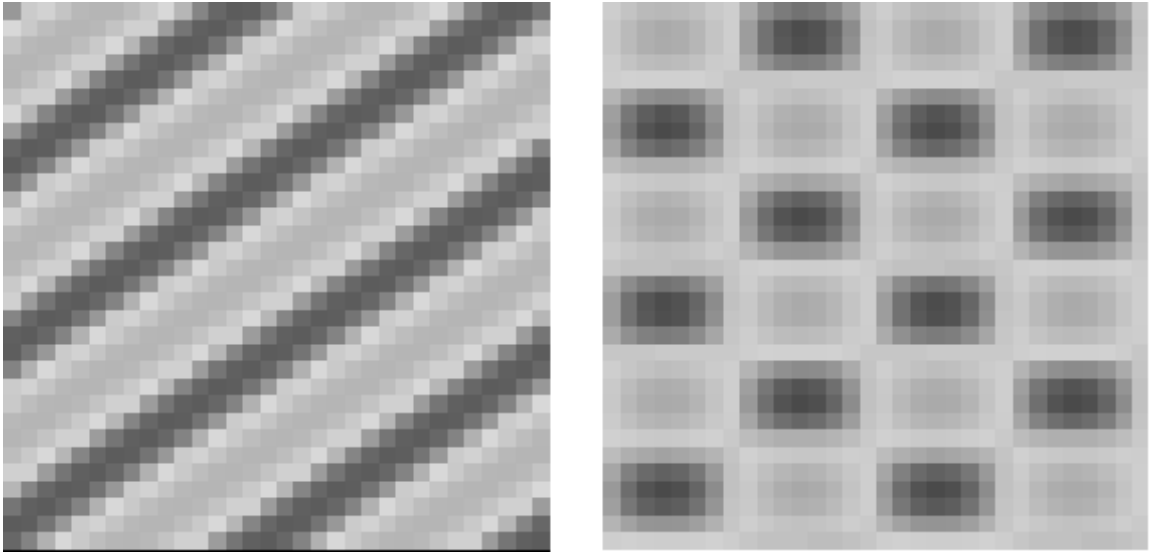} \caption{Rolls and
  Rhombs in Lengyel-Epstein CIMA model~(\protect\ref{eq:cimablah}) for
  $\delta=1$, $\hat b=5$, and $\hat c$ chosen close to the Turing
  bifurcation point, as discussed in the text.  Plot is of the
  numerically computed amplitude of the activator variable $u$,
  starting with a random initial condition.  Rolls were generated at
  $\hat a=9.74$, rhombs at $\hat a = 9.8$, in agreement with
  Figure~\ref{fig:cimaparms}. The computational domain corresponds to
  a (3,2)-square lattice; the grid size is $32\times 32$.}  
  \label{fig:cimarolls}
\end{figure}

\begin{figure}
  \epsfbox{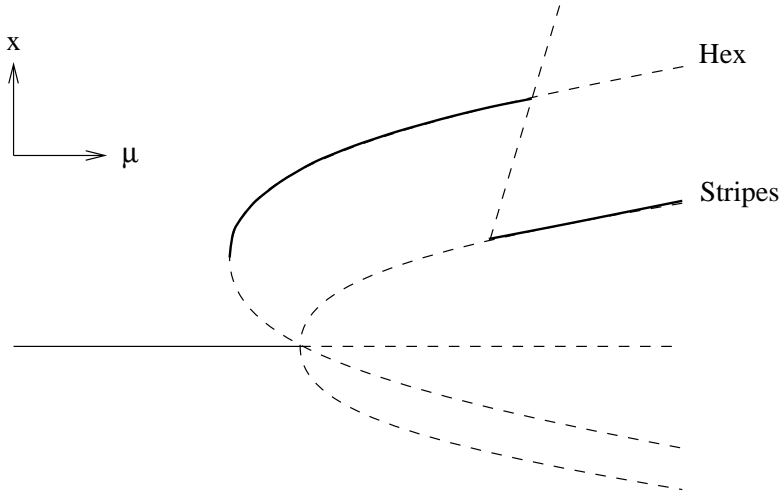}
  \caption{Unfolding of the degeneracy $(\hat a=9.39\ldots$) for the
Lengyel-Epstein CIMA model. All other primary solution branches
on the hexagonal lattices are unstable.}
  \label{fig:unfold}
\end{figure}


A number of general conclusions may be drawn from
Figure~\ref{fig:cimaparms}.  At all values of $\hat a$, simple squares
are unstable for some range of $\theta$-values.  We therefore conclude
that they are unstable, near onset, in the unbounded domain.  Rolls
are similarly unstable for all $\hat a$ with the exception of $\hat a
\approx 9.4$, where rolls are stable to perturbations in any direction
$\theta$. Below we discuss the significance of this particular
$\hat a$-value.

Figure~\ref{fig:cimarolls} shows numerical
simulations at two points in Figure~\ref{fig:cimaparms}.  A square
box size of $L=\sqrt{13} (2\pi/q_c)$ gives critical lattice vectors of 
$(\pm 3,2) (2\pi/L)$ and $(\pm 2,3) (2\pi/L)$ ({\it cf.} 
Figure~\ref{fig:lattice}).
A $(3,2)$-lattice generates mode angles of 
$\theta_3 = \cos^{-1} 12/13 \approx \pi/8$
and $\theta_4 = \pi/2 - \theta_3 \approx 3\pi/8$.
Figure~\ref{fig:cimaparms} suggests
that for $\theta = 3\pi/8$ rolls become unstable to rhombs at
$\hat a \approx 9.8$.  The rolls in Figure~\ref{fig:cimarolls} were
generated from random initial conditions at $\hat a = 9.74$; the
rhombs were generated from random initial conditions at $\hat a = 9.8$.
For these simulations $\delta = 1$, $\hat b = 5$, and the numerics
were performed quite close to the bifurcation point, using a value of
$K = K_c + 2.9\times 10^{-4}$.  These numerical results are in fact sensitive
to the size and shape of the computational domain; the choice of
the computational domain damps modes that might otherwise destabilize
the rhombs.

In Figure~\ref{fig:cimaparms}, rolls appear to become stable for all
$\theta$ at the point $\hat a \approx 9.4$.
This is the point at which the hexagonal
degeneracy condition $\gamma=0$ is satisfied;
since the nonlinear terms differ
by a constant, the only way to satisfy 
$\gamma \equiv \tilde u F_2(u_1, v_1) + \tilde v G_2(u_1,v_1) = 0$
in this model is to take $F_2(u_1,v_1) = G_2(u_1, v_1) = 0$,
which occurs at
$$
	x_0^2 = 2 + {\sqrt{21}\over 3} \Rightarrow \hat a = 9.39\ldots.
$$
At this point, hexagons are unstable to rolls ({\it cf.} 
Figure~\ref{fig:alphabet}
for $\alpha=0$, $\beta<0$).

The results summarized by Table~\ref{tab:evhex} may be used to
determine certain aspects of the  hexagonal
bifurcation problems near the degeneracy $\gamma=0$.  In
section~\ref{sec:hexdegen} it was pointed out that the unfolding
scenario is fairly simple; in this model, the unfolding computation is
even simpler.  Since $F_2(u_1,v_1) = 0$, the parameter $\alpha$ is
zero.  And since we know from section~\ref{sec:hexdegen} that there is
no $\theta$-dependence, the unfolding is described by the single
bifurcation diagram given in Figure~\ref{fig:unfold}.

%
%


\vskip 1cm
\subsection{An Example of Super Squares}\label{sec:supsq}


\begin{figure}
\centerline{
\epsfxsize=200pt\epsffile{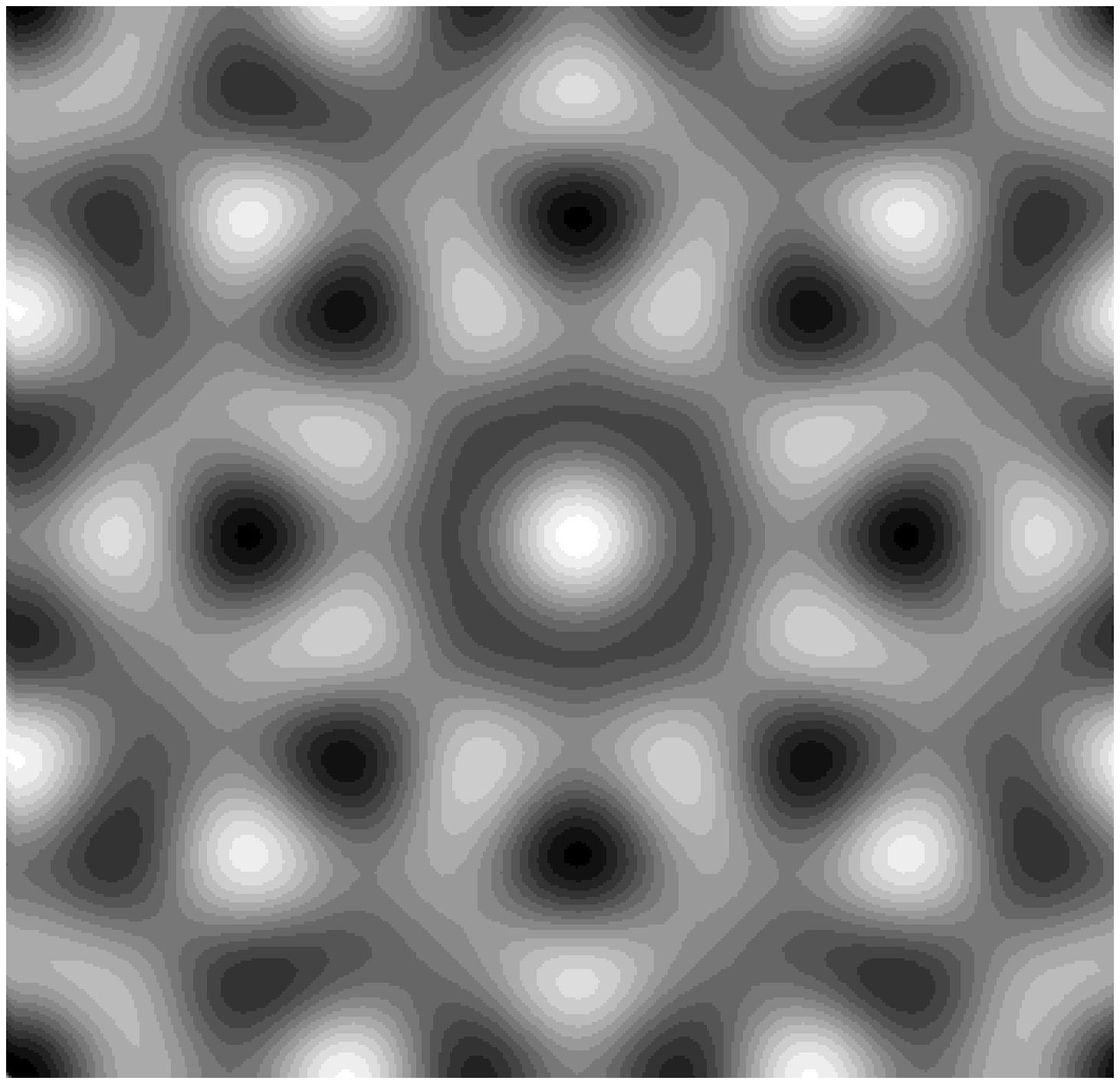}
\hskip 20pt
\epsfxsize=200pt
\epsffile{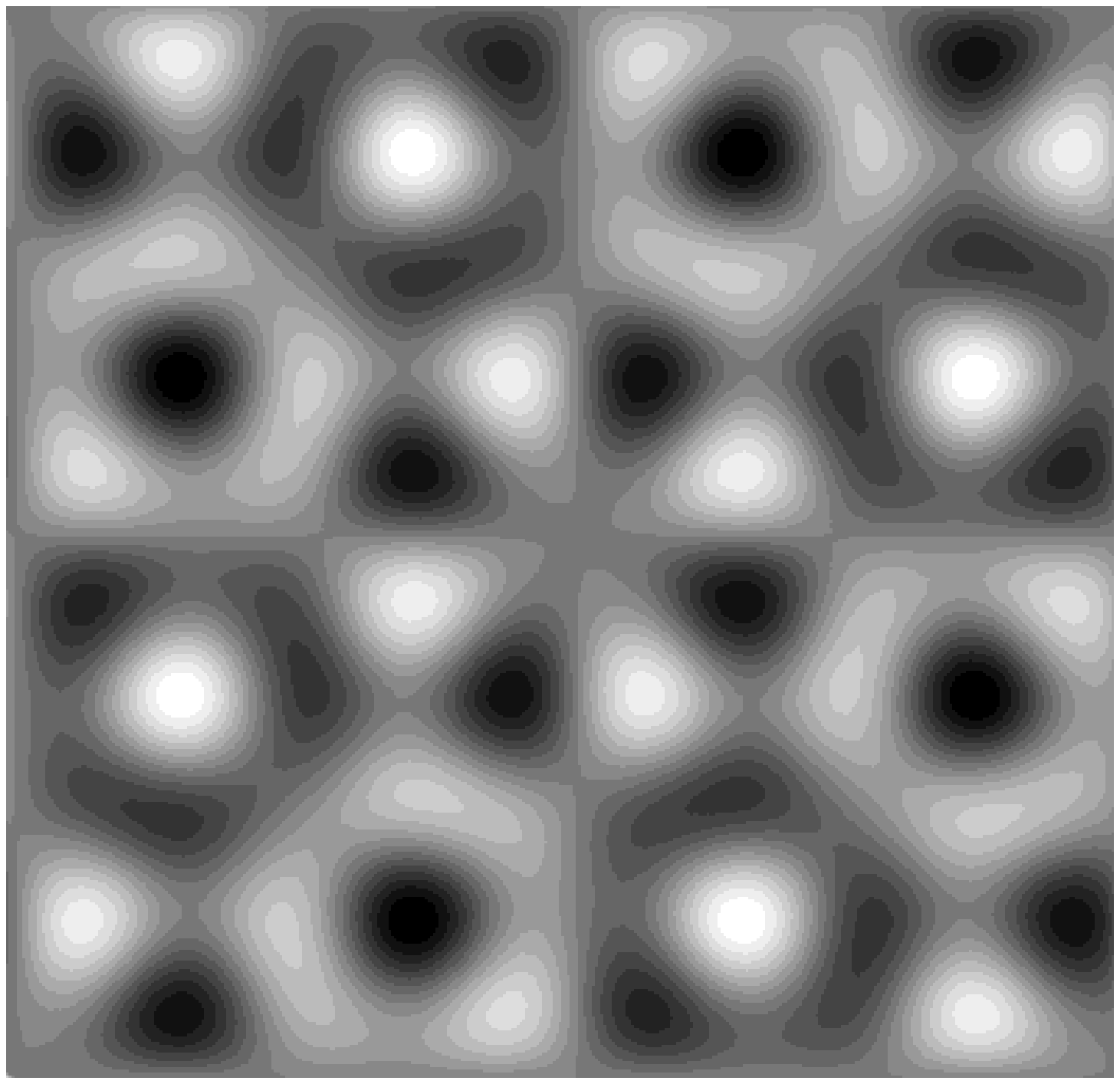}}
\caption{Examples of super squares and anti-squares for 
$(m,n)=(5,2)$,
\protect{\it i.e.} a box size of $\protect\sqrt{29}(2\pi/q_c)$.
Patterns consist of four equal-amplitude critical Fourier modes, as described
in section~\protect\ref{sec:formulation}.}
\label{fig:sasexample}
\end{figure}

Section~\ref{sec:formulation} described two types of superlattice
patterns for square lattices: super squares and anti-squares.
Figure~\ref{fig:sasexample} gives an example of each of these
patterns for $(m,n)=(5,2)$, corresponding to $\theta_3 = \cos^{-1}(20/29)$.
Each pattern has a $\pi/2$ rotational symmetry.  Super squares 
are invariant under reflections of the square domain; anti-squares
have symmetries involving both translations and reflections
(see~\cite{ref:mary} for details).

\begin{figure}
  \centerline{
    \epsfxsize=200pt\epsffile{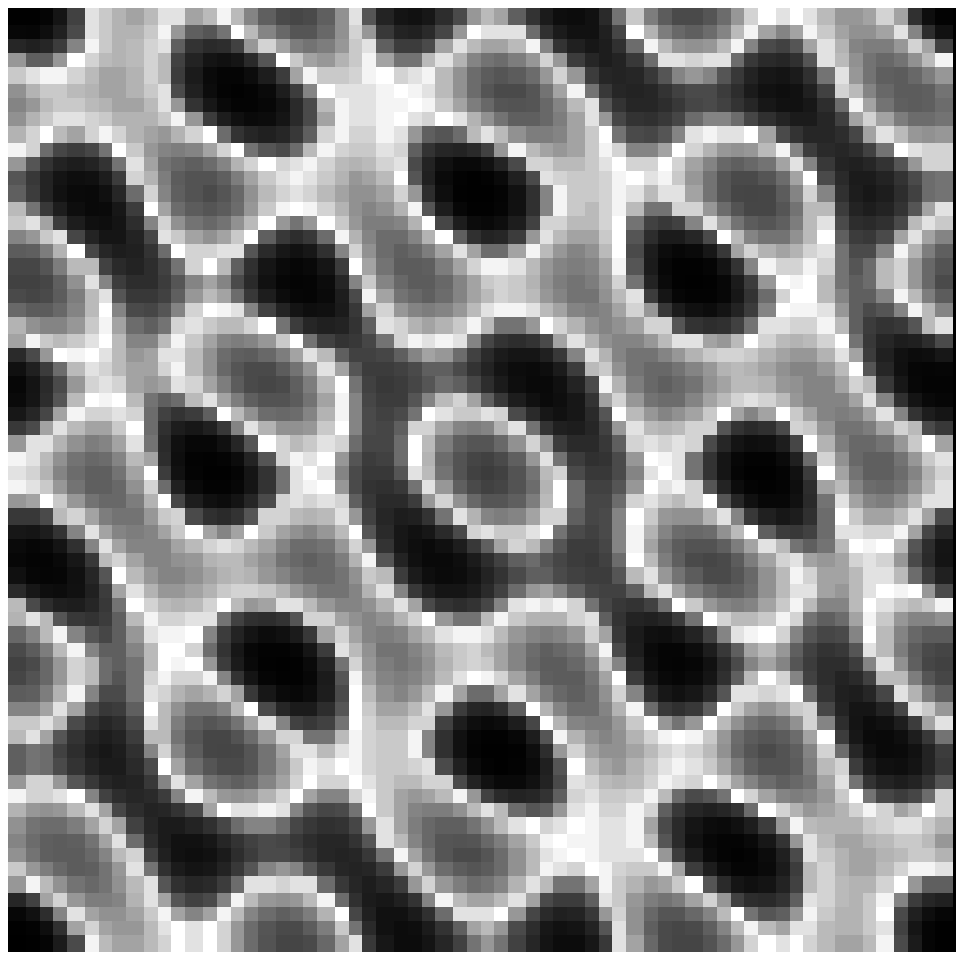}
    \hskip 25pt
    \epsfxsize=200pt
    \epsffile{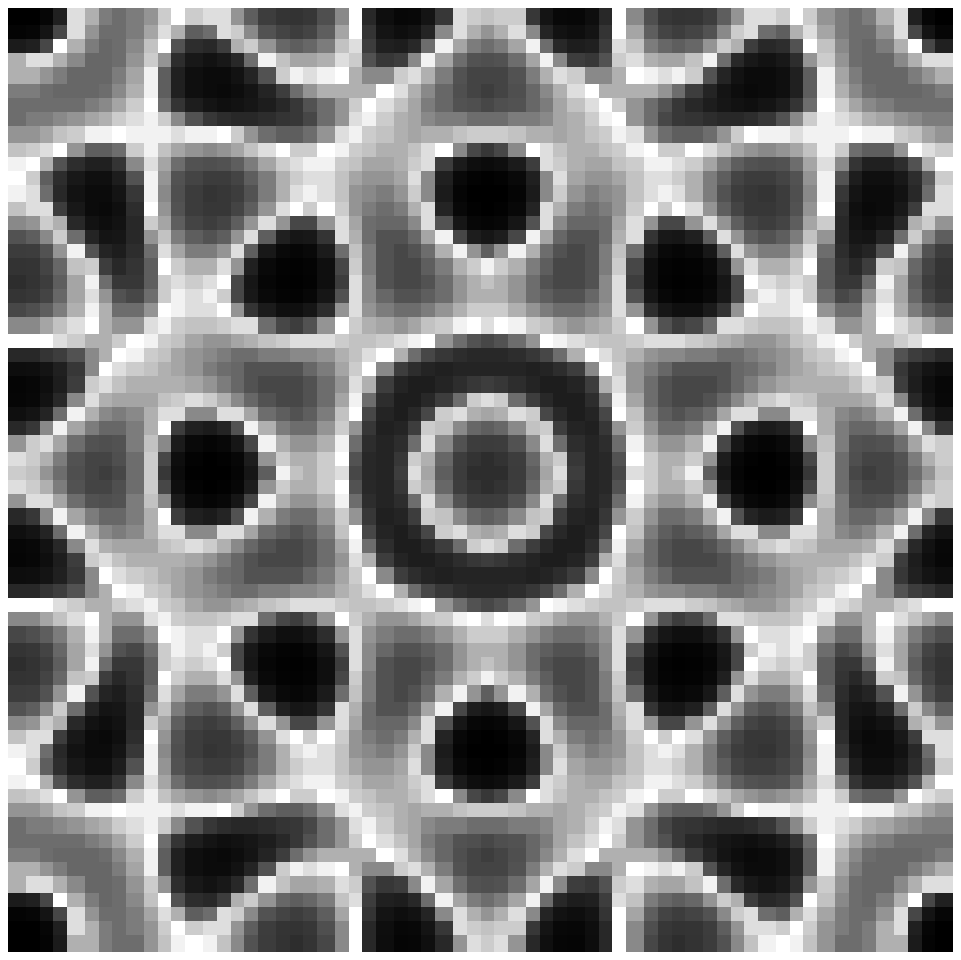}}

  \caption{Numerical integration of system
(\protect\ref{eq:supersquare}) 
described in Section~\protect\ref{sec:supsq}, with  a box of size 
$\protect\sqrt{29}(2\pi/q_c)$, and on a
	$64\times 64$ grid.
	The initial (left) and final (right) distributions of the
field $u({\bf r})$ are shown. Compare the final state  with the super squares of
 Figure~\protect\ref{fig:sasexample}.  }
  \label{fig:supsq}
\end{figure}

In Section~\ref{sec:collapse} a system was constructed which supported
stable super hexagons; by using the Appendix coefficient expressions, 
coupled with the stability information in Table~\ref{tab:evsq}, we may
similarly construct a system which supports either super or anti-squares.
(The relative stability of super and anti-squares is determined by
terms of higher than cubic order in~(\ref{eq:sqeq}), which we have not
computed.)  The reaction-diffusion 
system~(\ref{eq:full}) with
\begin{eqnarray}
\label{eq:supersquare}
f(u,v) &=& au - 1.8v - 0.15 uv + v^2,\nonumber\\
g(u,v)& =& 6u - 2.8v + 2.8v^2 + 0.2u^3,\\
K &=& 9,\nonumber
\end{eqnarray}
undergoes a Turing bifurcation at $a_c = 1.8797\ldots$ with
critical wave number $q_c = 0.88563\ldots$.  Figure~\ref{fig:supsq} 
shows a numerical integration of this system at $a=1.88$, just above
onset, showing both the initial condition and final steady state.

\section{Conclusions}

From the general two-component reaction-diffusion
system~(\ref{eq:full}), we have derived analytic expressions for the
coefficients of the leading nonlinear terms in the bifurcation
equations for rhombic, square and hexagonal lattices. These
coefficients allow us to calculate the relative stability of patterns
which are 
periodic on either square lattices (stripes, squares, rhombs, and
super squares or anti-squares) or hexagonal lattices (stripes, simple
hexagons, rhombs, super hexagons or super triangles) at the onset of the
Turing instability. In particular, the dependence of stability upon
system parameters may be calculated explicitly for
specific reaction-diffusion models,
as we demonstrated for several model equations,
including the Lengyel-Epstein CIMA model.


One of the surprising results of our general analysis is
that details of the reaction-kinetics enter the computation of the
bifurcation coefficients in a very limited fashion.  In
the case of the degenerate hexagonal bifurcation problem we show
that the coefficients of the cubic terms in the bifurcation problem
depend on the reaction-kinetics through just two effective parameters,
that are simple to compute. Moreover, we find that at this degeneracy
the angle dependence, which is expected for the rhombic lattice problems, 
drops out completely, and that rhombs and super
hexagons never bifurcate stably at the Turing bifurcation point of
two-component reaction-diffusion models.  However, in the unfolding 
of this degenerate bifurcation problem we find that super hexagons may
co-exist stably with simple hexagons and stripes for a limited range
of the bifurcation parameter. This tri-stability property is
demonstrated by numerical simulation of a reaction-diffusion system in
the vicinity of the degenerate bifurcation point.

We plan to carry
out more extensive numerical investigations that test some of the
predictions of our bifurcation analysis.  This will enable us to
investigate the domain of validity of our bifurcation analysis,
which is restricted to small-amplitude spatially-periodic Turing
patterns.  We also intend to determine the nature of the curious
behavior at the hexagonal degeneracy, and to determine to what
extent parameter collapse is exhibited by
systems with three or more components, {\it e.g.}
for Turing patterns in activator-inhibitor-immobilizer
models~\cite{ref:pearson}.

\section{Acknowledgments}
We have benefited from discussions with E. Knobloch, H. Riecke, and
A.C. Skeldon. The research of MS was supported by NSF grant
DMS-9404266 and by an NSF CAREER award DMS-9502266.


\newpage
\appendix
\section{Appendix}\label{ap:coeffs}

Here we summarize our calculations of the coefficients in
the bifurcation problems
associated with rhombs and hexagons, for systems of the form
\eqalign
   u_t &= &\nabla^2 u + au + bv + F_2(u,v) + F_3(u,v) + \cdots\\
   v_t &= &K\nabla^2 v + cu + dv + G_2(u,v) + G_3(u,v) + \cdots,
        \quad \nabla^2 = \partial_{xx} + \partial_{yy}.
\endalign
The calculation assumes that the linear coefficients are near the
Turing
bifurcation point, {\it i.e.},
$$a=a_c + \lambda_a, 
  \quad b=b_c + \lambda_b, 
  \quad c=c_c+\lambda_c, 
  \quad d=d_c + \lambda_d,
  \quad K=K_c + \lambda_K,
$$
$$(K_c a_c -d_c )^2 + 4K_c b_c c_c  = 0,$$
and that the system is in the Turing regime (conditions~(\ref{eq:negcond})).
The following values are common to all coefficient expressions: 
the right and left null vectors from the linear problem,
\eqalign
  \left(\matrix{u_1\cr v_1\cr}\right) \equiv 
	\left(\matrix{-b_c \cr a_c  - q_c^2\cr}\right), \quad
  \left(\matrix{\tilde u\cr \tilde v\cr}\right) \equiv
	\left(\matrix{c_c\cr -v_1\cr}\right);
\endalign
the critical wave number $q_c$, where
\eqalign
	q_c^2 \equiv {K_c a_c +d_c \over 2K_c };
\endalign
the quadratic and cubic nonlinearities from the Taylor expansion of
the reaction terms,
\eqalign
  \left(\matrix{F_2(u,v)\cr G_2(u,v)\cr}\right) &\equiv
	&\left(\matrix{f_{uu} u^2/2 + f_{uv}u v + f_{vv}v^2/2\cr
		g_{uu} u^2/2 + g_{uv} u v + g_{vv} v^2/2\cr}\right)\\
\medskip
  \left(\matrix{F_3(u,v)\cr G_3(u,v)\cr}\right) & \equiv
	&\left(\matrix{f_{uuu}u^3/6 + f_{uuv}u^2 v/2 + 
		f_{uvv} u v^2/2 + f_{vvv} v^3/6\cr
		g_{uuu}u^3/6 + g_{uuv}u^2 v/2 + g_{uvv}u v^2/2
		+ g_{vvv} v^3/6\cr}\right);
\endalign
and three effective parameters involving the quadratic and cubic 
nonlinearities,
\eqalign
\eta_1 &\equiv &\tilde u {\partial F_2(u_1,v_1)\over \partial u_1}
	+ \tilde v {\partial G_2(u_1, v_1)\over \partial u_1}\\
\eta_2 &\equiv &\tilde u {\partial F_2(u_1,v_1)\over \partial v_1}
	+ \tilde v {\partial G_2(u_1, v_1)\over \partial v_1}\\
\beta &\equiv &\tilde u F_3(u_1,v_1) + \tilde v G_3(u_1, v_1).
\endalign
The value of the linear term $\mu$ in the amplitude equations below 
is given by
$$\mu \equiv (\tilde u,\tilde v)\cdot \left(\matrix{\lambda_a &\lambda_b\cr
	\lambda_c &\lambda_d - q_c^2\lambda_K\cr}\right)
	\left(\matrix{u_1\cr v_1\cr}\right).
$$
Finally, an overall factor of $u_1 \tilde u + v_1\tilde v$, which 
is always positive,
has been removed from the equations by rescaling time.

\medskip
\subsection{Rhombic lattice}

The amplitude equation for the rhombic lattice problem is
\eqalign
  \dot z_1 &= &\mu z_1 + a_1 |z_1|^2 z_1 + h(\theta) |z_\theta|^2 z_1,\cr
  \dot z_\theta &= &\mu z_\theta + a_1 |z_\theta|^2 z_\theta + 
	h(\theta) |z_1|^2 z_\theta.
\endalign
The coefficients are given by
$$
  a_1= (u_2+u_5)\eta_1 + (v_2+v_5)\eta_2 + 3\beta,
%
$$
$$
  h(\theta) = (u_3+u_4+u_5)\eta_1 + (v_3+v_4+v_5)\eta_2 + 6\beta,
$$
\smallskip
where
$$
\left(\matrix{u_2\cr v_2\cr}\right) = {1\over 9K_c q_c^4}
	\left(\matrix{4K_c q_c^2-d_c  &b_c \cr c_c  &4q_c^2-a_c \cr}\right)
	\left(\matrix{F_2(u_1, v_1)\cr G_2(u_1, v_1)\cr}\right) ,
$$
$$
\left(\matrix{u_3\cr v_3\cr}\right) = 
	{2\over \xi_+}
	\left(\matrix{K_c (a_c  + 2q_c^2\cos\theta) &b_c \cr 
		c_c  &{d_c \over K_c } + 2q_c^2\cos\theta\cr}\right)
	\left(\matrix{F_2(u_1, v_1)\cr G_2(u_1, v_1)\cr}\right) ,
$$
$$
\left(\matrix{u_4\cr v_4\cr}\right) = 
	{2\over \xi_-}
	\left(\matrix{K_c (a_c  - 2q_c^2\cos\theta) &b_c \cr 
		c_c  &{d_c \over K_c } - 2q_c^2\cos\theta\cr}\right)
	\left(\matrix{F_2(u_1, v_1)\cr G_2(u_1, v_1)\cr}\right) ,
$$
$$
\left(\matrix{u_5\cr v_5\cr}\right) = 
	{2\over K_c q_c^4}
	\left(\matrix{-d_c  &b_c \cr c_c  &-a_c \cr}\right)
	\left(\matrix{F_2(u_1, v_1)\cr G_2(u_1, v_1)\cr}\right) ,
$$
with
\eq
  \xi_\pm = K_c q_c^4 (1\pm 2\cos\theta)^2.
\endeq
It may be shown that
\eqalign
  u_3 + u_4 + u_5 &= &{2\over K_c q_c^4 (1 - 4\cos^2\theta)^2}\biggl(
	F_2(u_1, v_1) \left(2 K_c a_c - d_c (1+ 16\cos^4\theta)\right) \cr
	&&\qquad\qquad\qquad\qquad \quad
	+ G_2(u_1, v_1) b_c (3 + 16 \cos^4\theta)\biggr),\\
\medskip
  v_3 + v_4 + v_5 &= &{2\over K_c q_c^4 (1 - 4\cos^2\theta)^2}\biggl(
	F_2(u_1, v_1) c_c  (3 + 16 \cos^4\theta) \cr
	&&\qquad\qquad\qquad\qquad \quad
	+ G_2(u_1,v_1) \left( 2{d_c\over K_c} - a_c (1+16\cos^4\theta)
	\right)\biggr).
\endalign

\medskip
\subsection{Hexagonal lattice}

The amplitude equation for the hexagonal lattice problem is
$$
  \dot z_1 = \mu z_1 + \gamma z_2^*z_3^*  + b_1 |z_1|^2 z_1 
	+ b_2 (|z_2|^2 + |z_3|^2)z_1,
$$
with $\dot z_2$ and $\dot z_3$ obtained by cyclically permuting 
$z_1, z_2, z_3$.
The coefficients are given by
$$
  \gamma = 2{\left(\tilde u F_2(u_1,v_1) +  \tilde v G_2(u_1,v_1)\right)
},
$$
$$
  b_1= a_1,
$$
$$
  b_2= (u_5 + u_6)\eta_1 + (v_5+v_6+v_7)\eta_2 - 
	{\gamma \tilde v v_7\over u_1\tilde u + v_1\tilde v}
	+ 6\beta,
$$
\smallskip
where $u_1$, $v_1$, $u_5$, and $v_5$ are given in the rhombic case, and
$$
\left(\matrix{u_6\cr v_6}\right) = 
	{1\over 2 K_c q_c^4}
	\left(\matrix{ 3 K_c q_c^2-d_c &b_c\cr c_c &3q_c^2-a_c\cr}\right)
	\left(\matrix{F_2(u_1, v_1)\cr G_2(u_1, v_1)\cr}\right) ,
$$
$$
	v_7 = {1\over b_c}\left({\gamma u_1\over u_1\tilde u + v_1\tilde v}
		 - 2F_2(u_1, v_1)\right).
$$







\end{document}